\documentclass[11pt,reqno,preprint]{article}
\usepackage{jheppub}
\usepackage{amssymb,amsmath}
\usepackage{color}
\usepackage{hyperref}
\usepackage{graphicx}
\usepackage{bbold}
\usepackage{MnSymbol}
\usepackage{subcaption}
\usepackage{ytableau}
\usepackage{tikz}
\title{Counting states in a model of replica wormholes}
\author{Henry Maxfield}
\affiliation{Stanford Institute for Theoretical Physics, Stanford University, Stanford, CA 94305, USA}
\emailAdd{henrym@stanford.edu}

\newcommand{\hilb}{\mathcal{H}}	
\newcommand{\ZZ}{\mathbb{Z}}		
\newcommand{\CC}{\mathbb{C}}		
\newcommand{\id}{\mathbb{1}}

\DeclareMathOperator{\supp}{supp}

\newcommand{\partition}{\vdash}

\usepackage{mathrsfs}

\DeclareMathOperator{\Tr}{Tr}
\DeclareMathOperator{\rank}{rank}
\DeclareMathOperator{\Sym}{Sym}

\graphicspath{{./Figures/}}

\abstract{We study the Hilbert space of a system of $n$ black holes with an inner product induced by replica wormholes. This takes the form of a sum over permutations, which we interpret in terms of a gauge symmetry. The resulting inner product is degenerate, with null states lying in representations corresponding to Young diagrams with too many rows. We count the remaining states in a large $n$ limit, which is governed by an emergent collective Coulomb gas description describing the shape of typical Young diagrams. This exhibits a third-order phase transition when the null states become numerous. We find that the dimension of the black hole Hilbert space accords with a microscopic interpretation of Bekenstein-Hawking entropy.
}

\begin{document}

\maketitle

\section{Introduction}

Perturbative quantum gravity has too many states. This puzzling fact underpins the black hole information problem \cite{Jacobson:2003vx,Mathur:2009hf,Harlow:2014yka,Unruh:2017uaw,Marolf:2017jkr}. A na\"ive approach to counting states in gravity is to choose some background spacetime and perturbatively quantise fluctuations of metric and matter fields; the $\log$ of the number of states arising in this way (while remaining within the regime of low-energy effective gravity and after accounting for constraints from diffeomorphism invariance) defines a microcanonical `perturbative entropy' $S_\mathrm{pert}$ for the background in question. On a black hole background $S_\mathrm{pert}$ scales with the volume of the black hole interior, which grows without bound as the black hole ages. This is in tension with the expected finite number of internal states of a black hole, which follows from a conventional microscopic interpretation of the Bekenstein-Hawking entropy $S_\mathrm{BH}\sim \frac{A}{4G_N}$. For a sufficiently old black hole $S_\mathrm{pert}\gg S_\mathrm{BH}$, so perturbative quantum gravity supplies an excess of states.

One possible solution to this puzzle is that gravitational effective field theory is not applicable for the interior of an old black hole, and so the states obtained in perturbation theory cannot be trusted. But since spacetime curvatures remain small in the region where we are applying the theory, there is no obvious reason why such a failure should occur.

An alternative is that the states obtained in perturbation theory are perfectly trustworthy, but are not in fact independent. Instead, they form an enormously redundant overcomplete  set of states in the non-perturbative physical Hilbert space. This means that there are many `null states': non-zero wavefunctions of perturbative states which have zero norm in the complete non-perturbative Hilbert space. Such redundancies might arise from finely-tuned non-perturbative corrections to the gravitational inner product.

More explicitly, let $|i\rangle$ be an orthonormal basis for a Hilbert space of states $\hilb_\mathrm{pert}$ in gravitational perturbation theory (perhaps consisting of sufficiently small fluctuations around some classical black hole background):
\begin{equation}\label{eq:Hpert}
	|i\rangle \in \hilb_{\mathrm{pert}},\quad \langle j|i\rangle = \delta_{ij}, \qquad i,j =1,2,\ldots,e^{S_\mathrm{pert}}.
\end{equation}
In the physical Hilbert space of the complete theory, the inner product receives corrections from non-perturbative effects such as spacetimes with different topologies (wormholes). We denote this physical inner product with double angle brackets $\llangle \cdot |\cdot \rrangle$, writing
\begin{equation}\label{eq:Hphys}
	|i\rrangle\in\hilb_\mathrm{phys},\qquad 
	\llangle j|i\rrangle = \langle j|\eta|i\rangle = \eta_{ji} = \delta_{ji}+R_{ji} \,,
\end{equation}
where $\eta$ is a Hermitian positive semi-definite operator on $\hilb_\mathrm{pert}$, with matrix elements $\eta_{ji}$. This matrix of inner products is close to the identity, in the sense that the corrections $R_{ji}=\eta_{ji}-\delta_{ji}$ to individual matrix elements are typically exponentially small. Nonetheless, such corrections can have a significant effect, enormously reducing the dimension of the physical Hilbert space:
\begin{equation}
	e^{S_\mathrm{phys}} =\dim \hilb_{\mathrm{phys}} = \rank \eta \ll  \dim \hilb_{\mathrm{pert}} = e^{S_\mathrm{pert}}.
\end{equation}
This can be true even if matrix elements $R_{ji}$ are typically only of order $e^{-\frac{1}{2}S_\mathrm{phys}}$, though they must be finely tuned so that $\eta$ is non-negative, but not of full rank. `Null states' $|N\rangle = \sum_i c_i |i\rangle \in\hilb_\mathrm{pert}$ are perturbative states in the kernel of $\eta$: $\sum_i \eta_{ji}c_i = 0$, so that the corresponding physical state has vanishing norm and hence must be identified with the zero state: $|N\rrangle = \sum_i c_i |i\rrangle = 0\in\hilb_\mathrm{phys}$ (a precise definition of null states and $\hilb_\mathrm{phys}$ is in section \ref{sec:IPdef}).

The purpose of this paper is to explicitly realise this idea in a simple model of black holes, where the relevant corrections arise from spacetime wormholes. In this model we give an explicit expression for $\eta$, identify precisely which states are null, and count the remaining physical states. Perhaps the most intriguing aspect is not the final result itself, but the structure which appears along the way. The state counting requires solution of an emergent Coulomb gas problem, arising from the representation theory of Young diagrams. This might be interpreted as an emergent collective `superspace field theory'\footnote{We mean Wheeler's notion of superspace  \cite{Misner:1973prb}, namely the space of configurations of spatial geometry, not to be confused with the unrelated supersymmetric notion.} which appears in the regime where null states become important.

While the model we study is rather simple, replica wormholes should contribute to the inner product in essentially the same way for any theory of gravity, as explained in section \ref{ssec:realistic}. (The main simplification of our model is that we have enough control to know that these are the only contributions, while we expect more realistic theories of gravity to have more complicated additional terms.) We thus anticipate the structure we encounter to be indicative of a universal mechanism underlying the counting of states in gravitational Hilbert spaces.

\subsection{The Page curve vs state counting}\label{sec:Page}

Before diving in, we make some conceptual points to help interpret our results. The first is to distinguish the notion of `state counting' in this paper from other recent work which probes the same physics, most prominently the gravitational calculation of the Page curve.

Here, we count states very directly: we identify the degrees of freedom defining the states (using gravitational variables) and find the inner product $\eta$, before diagonalising $\eta$ to count the dimension of the physical Hilbert space. In particular, we identify precisely which wavefunctions correspond to null states using a single non-random inner product.

In contrast, recent calculations of the Page curve and related work provide an indirect probe of the number of states. For this, create an entangled (pure) state of a black hole with a non-gravitational system, trace over the black hole, and compute the von Neumann entropy $S_\mathrm{vN}$ (or R\'enyi  entropies $S_n$) of the resulting state. This is related to the number of states $e^{S_\mathrm{phys}}$ since (assuming ordinary unitary quantum mechanics applies) it bounds the von Neumann entropy, $S_\mathrm{vN}\leq S_\mathrm{phys}$. Such an entangled state arises naturally between a black hole and its Hawking radiation as it evaporates, in which case the entropy of the radiation as a function of time should follow the `Page curve' \cite{Page:1993wv} if the number of physical states is indeed given by $e^{S_\mathrm{BH}}$.

This Page curve can be recovered from gravitational effective field theory \cite{Almheiri:2019psf,Penington:2019npb} by using the `quantum extremal surface formula' \cite{Engelhardt:2014gca} to compute $S_\mathrm{vN}$. This formula follows from certain non-perturbative gravitational effects, namely the contribution of `replica wormhole' spacetimes to the calculation of the entropy \cite{Penington:2019kki,Almheiri:2019qdq}. This provides strong quantitative evidence that the state-counting interpretation of the Bekenstein-Hawking formula ($S_\mathrm{phys}\sim S_\mathrm{BH}= \frac{A}{4G_N}$) is correct, but does not directly describe a physical Hilbert space of internal black hole states (such as constructing an inner product $\eta$ or identifying explicit wavefunctions of null states). Indeed, to interpret these results in terms of an inner product of black hole interior states appears to require some notion of statistics, with an ensemble of possible $\eta$s chosen from some classical probability distribution; the wormholes compute the statistics of this ensemble \cite{Coleman:1988cy,Giddings:1988cx,Marolf:2020xie,Saad:2018bqo,Penington:2019kki}.\footnote{This would perhaps be more usually stated as an ensemble for the density matrix of Hawking radiation, but these are essentially equivalent \cite{Marolf:2020rpm}. The perspective of directly computing statistics of inner products was taken in \cite{Balasubramanian:2022gmo}.}

While the Page curve results require non-perturbative effects, they essentially involve only a single wormhole configuration (or a family labelled by replica number).\footnote{A slight exception to this occurs if one wants to describe the details very close to the Page transition, where a larger set of wormholes should be included \cite{Marolf:2020vsi,Penington:2019kki}.} Our results require much more than this, with null states appearing only after summing over an exponentially large set of wormhole configurations. As such, the required effects are not even visible at any order in an expansion in the exponentially small parameter $e^{-S}$; they are `doubly non-perturbative'. This is similar to the physics required to resolve discreteness of the black hole spectrum (such as the `plateau' in the spectral form factor).

\subsection{The Hilbert space of $n$ black holes}\label{sec:HSn}

At this point, the reader may be surprised (and perhaps skeptical) that it is possible to explicitly identify null states in a black hole Hilbert space, since we would expect their wavefunctions to be extremely complicated and fine-tuned. We are able to do this because we study models not of a single black hole, but of many black holes with a non-factorising inner product. In this context, the null states can be more universal and less complicated. Here we briefly recall the main ideas underlying such models and the connection to the ensembles or randomness mentioned to above. We will not require any of the details in the remainder of the paper except for interpretation of the results (primarily \eqref{eq:interpent}).

The model we study describes the interior states of $n$ separate black holes (either very distantly separated, or perhaps in different universes entirely), each of which has $p$ internal states in perturbation theory around a particular background. So, the perturbative Hilbert space can be written as the $n$-fold tensor product
\begin{equation}\label{eq:HpertCpn}
	\hilb_\mathrm{pert} = (\CC^p)^{\otimes n}
\end{equation}
with the canonical inner product on $\CC^p$, giving a $p^n$-dimensional space or a total perturbative entropy $S_\mathrm{pert}=n\log p$. Each black hole will have a Bekenstein-Hawking entropy $S_\mathrm{BH} = \log q$, so in the full non-perturbative theory we would expect the dimension of the Hilbert space describing all states of all $n$ black holes to scale as $q^n$ (a total entropy $S_\mathrm{phys} \sim n S_\mathrm{BH} = n\log q$).

As in \eqref{eq:Hphys}, the physical Hilbert space $\hilb_\mathrm{phys} = \hilb^{(n)}$ is described by a matrix of inner products $\eta^{(n)}$ acting on $\hilb_\mathrm{pert}$. Ordinary local physics might lead one to expect that $\eta^{(n)}$ should act independently on each black hole, respecting the tensor product factorisation in \eqref{eq:HpertCpn} as $\eta^{(n)} = \eta^{(1)}\otimes\cdots \otimes \eta^{(1)}$ so that the Hilbert space $\hilb^{(n)} = \hilb^{(1)}\otimes\cdots\otimes \hilb^{(1)}$ similarly factorises. But a direct interpretation of spacetime wormholes violates this factorisation.

For an outside observer (e.g., for experiments performed on Hawking radiation) no observable violation of locality arises from this. Instead, the failure of factorisation can be accounted for by classical statistical correlations \cite{Coleman:1988cy,Giddings:1988cx,Marolf:2020xie}. In the context of AdS/CFT this outside perspective corresponds to the dual CFT, and gives rise to the idea of `ensemble duality' made famous by the example of JT gravity \cite{Saad:2019lba}. This is the statistical perspective alluded to in section \ref{sec:Page}. In terms of the Hilbert space, we have a weaker notion of factorisation: $\hilb^{(n)}$ decomposes into a direct sum of superselection sectors (labelled by some parameters $\alpha$), and the Hilbert space factorises within each sector:\footnote{We write $\subset$ instead of $=$ here because the full $n$-boundary Hilbert space describes closed `baby' universes in addition to the $n$ black holes in $\hilb^{(n)}$ \cite{Marolf:2020rpm}. Adding these baby universes changes the weighting of the state between different $\alpha$ sectors, or the distribution for the random inner product on the right-hand-side of \eqref{eq:IPvsEnsemble}.}
\begin{equation}\label{eq:superselection}
	 \hilb^{(n)} \subset \bigoplus_\alpha \underbrace{\hilb_\alpha \otimes \cdots \otimes \hilb_\alpha}_{n},
\end{equation}
which follows from a similar decomposition of $\eta$. Observables accessible asymptotically such as the black hole Hamiltonian are superselected, meaning that they respect this decomposition (being block diagonal in the $\alpha$ basis and factorising into the $n$ components in each block). This means that any distant experiments (on $n$ sets of Hawking radiation, for example) will be compatible with a factorising Hilbert space $\hilb_\alpha \otimes \cdots \otimes \hilb_\alpha$ for some fixed but randomly determined $\alpha$, and the asymptotic observer can use $\hilb_\alpha$ to describe the internal states of any single black hole.

The upshot is that the boundary (or asymptotic) interpretation of  $\eta^{(n)}$ naturally takes the form of an average:
\begin{equation}\label{eq:IPvsEnsemble}
	\llangle j_1,\ldots,j_n|i_1,\ldots,i_n\rrangle = \eta^{(n)}_{j_1,\ldots,j_n;i_1,\ldots,i_n}=  \overline{\langle j_1|i_1\rangle\cdots \langle j_n|i_n\rangle},
\end{equation}
where $\langle j|i\rangle$ denotes a random matrix of inner products and $\overline{\phantom{i}\cdots \phantom{i}}$ is an expectation value in some distribution. However, we are not forced to take this statistical or ensemble perspective, and for physics deep in the bulk (such as the interior of black holes which we are trying to describe) it is not even applicable as the relevant observables need not be superselected (since the superselection arguments of \cite{Marolf:2020xie} apply only to asymptotic operators). The correct gravitational interpretation in a theory with spacetime wormholes is really the single inner product on the left hand side, not an average.

The ensemble interpretation of $\eta^{(n)}$ is nevertheless useful to understand how the dimension of $\hilb^{(n)}$ should behave. A microscopic interpretation of Bekenstein-Hawking entropy requires that $S_\mathrm{BH}$ counts the internal states as deduced by an outside observer. This means the ensemble interpretation is appropriate, and that $S_\mathrm{BH}$ bounds $\log \dim \hilb_\alpha$, the internal states of any single black hole in a typical $\alpha$-sector. We therefore expect $\log \dim \hilb_\alpha \sim \min\{S_\mathrm{pert},S_\mathrm{BH}\}$, and the microscopic entropy for $n$ black holes will be $n$ times this. However, $\hilb^{(n)}$ encompasses states from many different $\alpha$-sectors so some of its entropy will be associated with the determination of the random variables $\alpha$. Nothwithstanding, we expect that this latter contribution of the entropy will grow more slowly if we take the number $n$ of black holes to be very large. As a result, we consider the large $n$ limit of the `entropy per black hole' $\frac{1}{n} \log \dim \hilb^{(n)}$, 
\begin{equation}\label{eq:interpent}
	 S=  \lim_{n\to \infty}\frac{1}{n} \log \dim \hilb^{(n)}  = \min\{S_\mathrm{pert},S_\mathrm{BH}\}.
\end{equation}
The second equality is our main result, matching this state-counting entropy in our model with the expectation from the Bekenstein-Hawking entropy.

For further heuristic motivation, imagine that there is some finite number $K$ of $\alpha$-sectors in the decomposition \eqref{eq:superselection}, and that each $\hilb_\alpha$ has roughly the same dimension $e^{S}$. Then the $\log$ of the dimension of $\oplus_\alpha \hilb_\alpha^{\otimes n}$ \eqref{eq:superselection} is $\log K + n S$; we interpret the first term as a classical entropy associated with uncertainty in $\alpha$. However, our model has infinitely many $\alpha$-sectors, so this rough argument does not apply and it is not obvious that the result in \eqref{eq:interpent} is necessary. We nonetheless find that it holds.

\subsection{Outline}

In section \ref{sec:model}, we define the Hilbert space and inner product we will be studying in the remainder of the paper. We motivate this inner product from gravity, first as arising from a sum over topologies and second from the perspective of a non-perturbative gauge symmetry.

With this inner product, in section \ref{sec:null} we identify the null states. To do this, we first characterise the inner product in terms of representation theory, decomposing the perturbative Hilbert space into sectors labelled by Young diagrams. The null states then have simple description: they are sectors corresponding to Young diagrams with too many rows.

We then begin the task of counting the null states in section \ref{sec:Coulomb}, introducing formulas for the dimensions of representations corresponding to given Young diagrams. These formulas have a physical interpretation in terms of the energy of a gas of particles with Coulombic interactions, and the dimension of the physical Hilbert space can be written as a classical statistical partition function of this Coulomb gas. By taking a continuum limit of this system, we arrive at an emergent collective description of many universes.

Using this collective description, in section \ref{sec:counting} we calculate the behaviour of the dimension of the physical Hilbert space remaining after removing the null states. In particular, we identify a third-order phase transition where the majority of states become null, and find the large $n$ limit given in \eqref{eq:interpent}.

We conclude in section \ref{sec:disc} with a discussion of further details of the inner product, generalisations, and interpretation of the results in more realistic models of gravity. Appendices contain details of JT gravity as an example of our inner product, more discussion of representation theory, some calculations, and the connection to a famous mathematical problem.

\section{The model} \label{sec:model}

In this section we introduce the Hilbert spaces $\hilb_{p,q}^{(n)}$ we are interested in. We will begin simply with a mathematical definition, with the inner product defined by a sum over permutations $\pi\in\Sym(n)$. We follow that with two (related) motivations for this definition from gravity, one geometric and one algebraic. The first comes from a sum over topologies, each permutation $\pi$ labelling a different topology of spacetime. The second motivation comes from a non-perturbative gravitational gauge symmetry, where $\Sym(n)$ forms part of the gauge group and the sum over permutations projects onto gauge-invariant states.

\subsection{The definition}\label{sec:IPdef}

We construct $\hilb_{p,q}^{(n)}$ beginning from $(\CC^p)^{\otimes n}$, the tensor product of $n$ copies of a $p$-dimensional vector space. We first equip $\CC^p$ with a canonical inner product with orthonormal basis $\{|i\rangle: i=1,2,\ldots,p\}$, so $(\CC^p)^{\otimes n}$ inherits the inner product
\begin{equation}
	\langle j_1,\ldots,j_n|i_1,\ldots,i_n\rangle = \delta_{i_1j_1}\cdots \delta_{i_nj_n} \,.
\end{equation}
We interpret this $p^n$-dimensional space as a Hilbert space constructed from perturbative gravity. We are interested in a Hilbert space which incorporates certain non-perturbative effects, which uses the same underlying vector space $(\CC^p)^{\otimes n}$ but a different inner product denoted with double angle brackets $\llangle \cdot |\cdot \rrangle$. We can write this `physical' inner product as
\begin{equation}
	\llangle \psi'|\psi\rrangle = \langle \psi'|\eta_{p,q}^{(n)} |\psi\rangle,
\end{equation}
where the matrix of inner products $\eta_{p,q}^{(n)}$ is a Hermitian operator on  $(\CC^p)^{\otimes n}$.

For this to define an inner product, we require that $\eta_{p,q}^{(n)}$ is positive semi-definite. It will be important for us to allow $\eta_{p,q}^{(n)}$ to have a nontrivial kernel, so some nonzero states in $(\CC^p)^{\otimes n}$ will have zero physical norm; these are `null states' $|N\rangle$ (by definition). Given this, we should interpret physical states as cosets in $(\CC^p)^{\otimes n}$, equivalence classes of states which differ by a null state:
\begin{equation}
	|\psi\rrangle = \{|\psi\rangle + |N\rangle :|N\rangle \in \ker \eta_{p,q}^{(n)} \} \in \hilb_{p,q}^{(n)}\simeq (\CC^p)^{\otimes n}/\ker \eta_{p,q}^{(n)}\,.
\end{equation}
Our main task will be to determine the dimension $\dim \hilb_{p,q}^{(n)}$, which is $p^n$ minus $\dim\ker \eta_{p,q}^{(n)}$, or equivalently the rank of $\eta_{p,q}^{(n)}$.

We define $\eta_{p,q}^{(n)}$ in terms of permutation operators acting on $(\CC^p)^{\otimes n}$. There is a natural unitary representation $R$ of the symmetric group $\Sym(n)$ acting on $(\CC^p)^{\otimes n}$ by permuting the $n$ factors:
\begin{equation}
	R(\pi) | i_1,\ldots,i_n\rangle = |i_{\pi^{-1}(1)},\ldots, i_{\pi^{-1}(n)}\rangle, \qquad \pi\in \Sym(n).
\end{equation}
Our matrix of inner products is simply a weighted sum over these permutations, depending on a parameter $q$:
\begin{equation}\label{eq:etapqn}
	\eta_{p,q}^{(n)} = \sum_{\pi\in\Sym(n)} q^{-|\pi|}\, R(\pi).
\end{equation}
The coefficients are written in terms of a function $|\pi|$ of $\Sym(n)$, which measures the `size' of a permutation. It is defined as $n$ minus the number of cycles in $\pi$. The minimal size is the identity $\id$ (a product of $n$ trivial one-cycles) with $|\id|=0$, and the maximum is a cyclic permutation like $\pi = (12\cdots n)$ for which $|\pi|=n-1$. Alternatively, $|\pi|$ is the minimal number of terms required to write $\pi$ as a product of transpositions. More explicitly we can also write the inner product in terms of the basis states $|i_1,\ldots,i_n\rrangle$, as
\begin{equation}\label{eq:IPdef}
	\llangle j_1,\ldots,j_n|i_1,\ldots,i_n\rrangle =  \sum_{\pi\in\Sym(n)} q^{-|\pi|} \delta_{i_1  j_{\pi(1)}}\cdots \delta_{i_n  j_{\pi(n)}} \,.
\end{equation}

As stated in the introduction after \eqref{eq:HpertCpn}, we will interpret $q= e^{S_\mathrm{BH}}$ as the exponential of the Bekenstein-Hawking entropy. If $S_\mathrm{BH}$ has some state-counting interpretation, it is natural to take $q$ to be a positive integer. This is in fact required for $\eta_{p,q}^{(n)}$  to be positive semi-definite for all $n$; see comments in section \ref{sec:disc}. This restriction will be further motivated in the following subsections.

\subsection{Motivation 1: a sum over topologies}\label{ssec:topmot}

Consider a simple two-dimensional model for a black hole, where the degrees of freedom in the interior are modelled by the state of an end-of-the-world (EOW) brane \cite{Penington:2019kki}. We can take this to be JT gravity as in  \cite{Penington:2019kki} or the even simpler topological model introduced in \cite{Marolf:2020xie} (only the topologies will ultimately be important for us, not the details of the dynamics.).  We take an EOW brane to have a $p$-dimensional space of internal states with orthonormal basis labelled by $i=1,2,\ldots,p$.

The spatial slice of a black hole is a line segment bounded by an asymptotic boundary (asymptotically AdS, for example) and an EOW brane. The wavefunction of the internal states of the EOW brane gives us a $p$-dimensional Hilbert space, so $n$ such black holes have a $p^n$-dimensional Hilbert space of states spanned by $|i_1,\ldots,i_n\rangle$ if we ignore topology change.

We would like to understand the effect of summing over spacetime topologies on the inner product of the states of these $n$ black holes. We can define $p^n$ states $|i_1,\ldots,i_n\rrangle$ non-perturbatively through boundary conditions: we have $n$ distinguishable asymptotic boundaries, each of which is required to meet an EOW brane with the specified label. For the inner product $\llangle j_1,\ldots,j_n|i_1,\ldots,i_n\rrangle$ we glue two such states, to obtain $n$ (oriented) line segments of asymptotic boundary which begin and end at EOW branes with labels $i_r$ and  $j_r$ respectively, with $r=1,\ldots,n$. To calculate the inner product we sum over all topologies with such boundary conditions.

\begin{figure}
\centering

\begin{tikzpicture}
	
	\node at (-7,0){\Large $\llangle j_1,\ldots,j_6|i_1,\ldots,i_6\rrangle \supset $};
	\node at (-3.4,-.3){\includegraphics{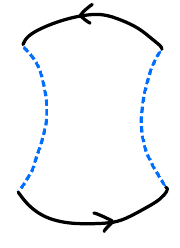}};
	\node at (1,0){\includegraphics{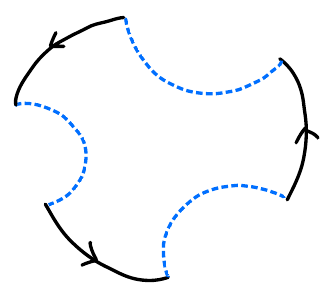}};
	\node at (5,0){\includegraphics{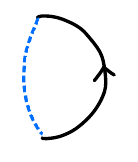}};	
	\node at (-4.2,-1.4){\Large $i_1$};
	\node at (-2.4,-1.4){\Large $j_1$};
	\node at (-2.5,.9){\Large $i_4$};
	\node at (-4.1,1){\Large $j_4$};
	\node at (-.6,-1.1){\Large $i_2$};
	\node at (.7,-1.9){\Large $j_2$};
	\node at (3.,-.4){\Large $i_6$};
	\node at (3,1){\Large $j_6$};
	\node at (.2,1.9){\Large $i_3$};
	\node at (-1.1,1.1){\Large $j_3$};
	\node at (4.7,-.7){\Large $i_5$};
	\node at (4.7,.8){\Large $j_5$};
\end{tikzpicture}
    \caption{A contribution to the inner product  \eqref{eq:EOWIP} between EOW brane states for $n=6$. The black curves correspond to segments of (oriented) asymptotically AdS boundaries, which must begin and end at EOW branes with specified species $i,j$. The blue dashed curves are EOW brane boundaries, which are dynamical meaning that we sum over all ways in which they join the asymptotic boundaries. The possibilities are labelled by a permutation $\pi$, in this case $\pi=(14)(263)(5)$, and we get a nonzero contribution to $\llangle j_1,\ldots,j_6|i_1,\ldots,i_6\rrangle$ only when $j_1=i_4$, $i_1=j_4$ and so forth. We have pictured the leading order topology for this $\pi$, which consists of three disconnected discs (one for each cycle); the topological action weights this geometry by  $e^{3S_0}$, corresponding to $\chi=n-|\pi|=3$. Including more complicated topologies with the same boundaries (along with the energy dynamics) only has the effect of replacing $e^{S_0}$ with the integer $q$.}
    \label{fig:JT}
\end{figure}
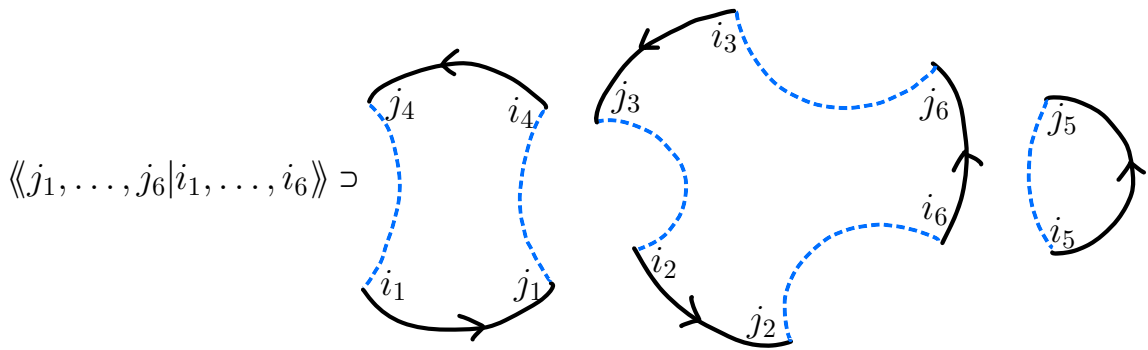

The EOW branes are dynamical, which means that we sum over all their configurations compatible with the boundary conditions. This means that we sum over the $n!$ different ways in which they connect the $n$ asymptotic boundary segments. In a given topology, the EOW brane beginning on boundary $r$ can end on boundary $\pi(r)$, where $\pi\in\Sym(n)$ is a permutation of the $n$ boundaries. This gives us a nonzero result only if the species match, so $i_r = j_{\pi(r)}$ for all $r=1,\ldots,n$. The contribution of a fixed topology to $\llangle j_1,\ldots,j_n|i_1,\ldots,i_n\rrangle$ is therefore proportional to a product of $n$ Kronecker deltas, $\delta_{i_1  j_{\pi(1)}}\cdots \delta_{i_n  j_{\pi(n)}}$.

Once we have decided how to connect up the EOW branes by choosing $\pi\in\Sym(n)$, we must fill in the remainder of the geometry. The boundary of the spacetime will consist of a number of circles, each consisting of alternating asymptotic boundary segments and EOW branes connected according to $\pi$. There is one such boundary for each disjoint cycle in $\pi$. So using the notation $|\pi|$ introduced above, the topology of spacetime will be a surface with $n-|\pi|$ boundaries.

The simplest such topology (which dominates since it has the smallest Euclidean action) is a union of discs; an example is shown in figure \ref{fig:JT}. To compute the contribution of this geometry, we start by using only the two-dimensional Einstein-Hilbert action, which is topological: the Euclidean action is $-S_0 \chi$, where $\chi$ is the Euler characteristic of spacetime. For a disc, we have $\chi=1$, so for our collection of discs we have action $-S_0(n-|\pi|)$. Summing over all permutations with this weighting gives us an inner product which includes topology change:
\begin{equation}\label{eq:EOWIP}
	\llangle j_1,\ldots,j_n|i_1,\ldots,i_n\rrangle = e^{n S_0 } \sum_{\pi\in\Sym(n)} e^{-S_0 |\pi|} \delta_{i_1  j_{\pi(1)}}\cdots \delta_{i_n  j_{\pi(n)}} \,.
\end{equation}
Up to the unimportant normalisation factor $e^{n S_0 }$ (which can be absorbed into the definition of the states), this is precisely the inner product \eqref{eq:IPdef}, with $p$ given by the number of EOW brane states and $q = e^{S_0}$.

There are two things we have left out: the more detailed dynamics (the JT action in the model of \cite{Penington:2019kki}) and more complicated topologies such as cylinders and higher genus surfaces. Typically, this will make the coefficients in \eqref{eq:EOWIP} more complicated, and introduce dependence on the temperature of each black hole. But there is a simple context in which we can include these effects and recover \eqref{eq:IPdef}, effectively replacing $e^{S_0}$ by a parameter $q$ which is constrained to be a positive integer. Specifically, we (1) choose `microcanonical' boundary conditions which constrain the energy of the states to lie in a narrow window, and (2) partially fix the spectrum of the JT Hamiltonian by specifying that a fixed number of energy eigenvalues lie in the chosen microcanonical window.\footnote{In other words, we fix some data about `which member of the ensemble' or which $\alpha$-state we work in, but only regarding the Hamiltonian, without saying anything about the wavefunctions of EOW brane states. Without any such restriction, the dimension of Hilbert space for large $n$ will be dominated by very rare $\alpha$-states in which the entropy fluctuates far from its semiclassical value.} In this case we arrive at the inner product \eqref{eq:IPdef} with $q$ giving the number of states in the chosen microcanonical window of energies. We explain this and discuss the inner product with the fixed temperature `canonical' boundary conditions in appendix \ref{app:JTIP}. Similarly, in the topological model of \cite{Marolf:2020xie} we recover \eqref{eq:IPdef} if we consider the inner product of EOW brane states in a sector of fixed $Z=q$. For more realistic theories of gravity, we expect a similar structure to appear from replica wormholes in fixed area states \cite{Akers:2018fow,Dong:2018seb} (of which a general state is a superposition); see section \ref{ssec:realistic}.

In \cite{Penington:2019kki}, the path integral we used to compute \eqref{eq:EOWIP} was given a slightly different interpretation in terms of an ensemble, as explained in section \ref{sec:HSn}. Namely, it was interpreted as giving the $n$th moments of a random inner product $\langle j|i\rangle$ as in \eqref{eq:IPvsEnsemble} (with the precise probability distribution described in appendix \ref{app:JTIP}). This is appropriate from the perspective of the boundary `dual' (namely an ensemble of quantum mechanical theories with a random Hamiltonian and random states $|i\rangle$), but the interpretation of $\llangle j_1,\ldots,j_n|i_1,\ldots,i_n\rrangle $ as a single definite inner product in the $n$-boundary Hilbert space is natural from the perspective of the bulk  theory of JT gravity.

\subsection{Motivation 2: gauging of non-perturbative diffeomorphisms}\label{ssec:gaugemot}

We have seen how the inner product \eqref{eq:IPdef} arises from a `path integral', summing over geometries of spacetime. Here we will see an alternative `canonical' derivation of the same inner product, coming from gauging a symmetry. This perspective will also give us a new way to write the inner product which will be technically useful later.

First we discuss this abstractly. Suppose we have a theory with finite group symmetry $G$, with a unitary representation $R$ acting on its Hilbert space $\hilb_\mathrm{kin}$: $R$ is a homomorphism mapping $G$ to the unitaries $U(\hilb_\mathrm{kin})$. We would like to gauge this symmetry to obtain a `physical' Hilbert space $\hilb_\mathrm{phys}$ from the `kinematic' ungauged space $\hilb_\mathrm{kin}$. For this we simply project onto $G$-invariant states of $\hilb_\mathrm{kin}$. One way to write the orthogonal projection $P$ onto this subspace is as a sum over the action of the group, $P = \frac{1}{|G|} \sum_{g\in G} R(g)$. Up to an unimportant normalisation, this means that we can write the physical inner product $\llangle \cdot|\cdot\rrangle$ in terms of the kinematic inner product $\langle \cdot|\cdot\rangle$ and the representation $R$ of $G$, as
\begin{equation}\label{eq:Gaverage}
	\llangle \psi'|\psi\rrangle = \sum_{g\in G} \langle \psi'|R(g)|\psi\rangle.
\end{equation}

From this perspective, we think of a physical state $|\psi\rrangle\in \hilb_\mathrm{phys}$ as a coset in $\hilb_\mathrm{kin}$ under the equivalences generated by $R(g)|\psi\rangle \sim |\psi\rangle$ for all $g\in G$. This coset is a `gauge orbit' of the state, and $R(g)|\psi\rangle \sim |\psi\rangle$ is a quantum gauge equivalence. Alternatively one might define $\hilb_\mathrm{phys}$ by `gauge-fixing', choosing a unique state in each coset such as the $G$-invariant $P|\psi\rangle$; this is equivalent (though subtleties may arise when considering instead a non-compact group).\footnote{Considering a Lie group $G$ instead, these alternatives correspond to `co-invariants' (equivalence classes of states differing by the image of the Lie algebra) and `invariants' (states annihilated by the Lie algebra). Under BRST quantisation these arise at the `top' and `bottom' level of cohomology respectively (i.e., maximal and minimal ghost number). There is a natural sesquilinear pairing between the invariants and co-invariants. The projection $P$ (or, generalising to non-compact $G$, a `group average') is a `rigging' map $\eta$ from co-invariants to invariants, which (along with the aforementioned pairing) defines an inner product on the co-invariants \cite{Marolf:2000iq} (an appendix of \cite{Chandrasekaran:2022cip} contains a useful introduction to these ideas).}

Now \eqref{eq:Gaverage} is similar to the matrix of inner products in \eqref{eq:etapqn} with $G=\Sym(n)$, but does not appear to allow for coefficients in the sum which depend on $g$. To explain where these might come from, we generalise to a situation with some additional unknown degrees of freedom: perhaps there are some short-distance modes that we have integrated out, for example. If these additional degrees of freedom are charged under the symmetry $G$, they will enter the physical inner product as $g$-dependent coefficients.

To make this more precise, suppose we have an extended kinematic Hilbert space which splits as $\hilb_\mathrm{kin}\otimes \tilde{\hilb}_\mathrm{kin}$. Take a representation of $G$ which similarly splits as a tensor product, acting as $R(g)\otimes \tilde{R}(g)$ for a pair of representations $R,\tilde{R}$ on $\hilb_\mathrm{kin}$, $\tilde{\hilb}_\mathrm{kin}$ respectively. We allow ourselves to freely choose the state on $\hilb_\mathrm{kin}$, but `integrate out' $\tilde{\hilb}_\mathrm{kin}$ by constraining ourselves to a fixed (mixed) state $\tilde{\rho}$ on that factor.\footnote{We can also always `purify' $\tilde{\rho}$ by including another additional factor on which $G$ acts trivially, and choosing a pure state which is entangled between $\tilde{\hilb}_\mathrm{kin}$ and.the extra factor. Then $\tilde{\rho}$ is the reduced density matrix of that state on $\tilde{\hilb}_\mathrm{kin}$.} Now we define a physical Hilbert space by summing over $G$ as in \eqref{eq:Gaverage}, but tracing out the additional factor in the chosen state, resulting in an inner product
\begin{equation}\label{eq:Gaverage2}
	\llangle \psi'|\psi\rrangle = \sum_{g\in G}  \widetilde{\Tr}\left(\tilde{\rho}\,\tilde{R}(g)\right) \langle \psi'|R(g)|\psi\rangle.
\end{equation}
We can also write this compactly in terms of the matrix of inner products $\eta$ on $\hilb_\mathrm{kin}$ as
\begin{equation}\label{eq:etaProj}
	\eta = |G| \,\widetilde{\Tr} \left(P \tilde{\rho}\right),
\end{equation}
where $P$ is the orthogonal projector onto $G$-invariant states of the product $\hilb_\mathrm{kin}\otimes \tilde{\hilb}_\mathrm{kin}$, and $\widetilde{\Tr}$ is the partial trace over the second factor.\footnote{An inner product of this form, though with an integral over a non-compact Lie group, appears naturally in recent considerations of de Sitter space \cite{Chandrasekaran:2022cip}. In that case, we take $\hilb_\mathrm{kin}$ to be the Hilbert space of QFT in dS and $\tilde{\hilb}_\mathrm{kin}$ the Hilbert space describing a clock, with $\tilde{\rho}$ some fixed initial state of the clock (a clock reading `midnight', say). Then, gauging the de Sitter isometries (which includes a Hamiltonian) by group averaging in the presence of the clock induces an inner product on $\hilb_\mathrm{kin}$, with states `dressed' to the clock.} In this form, the inner product is manifestly positive semi-definite (since $P$ is a positive semi-definite operator). To see this, by diagonalising $\tilde{\rho} = \sum p_k |k\rangle\langle k|$ (with $p_k>0$) we write $\eta$ as a sum of non-negative operators.

We can obtain our specific model \eqref{eq:etapqn} in this way by choosing the `hidden' degrees of freedom  $\tilde{\hilb}_\mathrm{kin} = (\CC^q)^{\otimes n}$  to take the same form as $\hilb_\mathrm{kin} = (\CC^p)^{\otimes n}$ with $\Sym(n)$ acting by permutations of the $n$ factors (both for $R$ and $\tilde{R}$). Additionally, we take $\tilde{\rho} = \tfrac{1}{q^n}\id$ to be the maximally mixed state on this space. This gives us
\begin{equation}
	\widetilde{\Tr}\left(\tilde{\rho}\,\tilde{R}(\pi)\right) = q^{-|\pi|} \qquad \left(\tilde{\hilb}_\mathrm{kin}=(\CC^q)^n,\quad \tilde{\rho} = \tfrac{1}{q^n}\id\right).
\end{equation}
To see this, note that the trace $\widetilde{\Tr}\left(\tilde{R}(\pi)\right)$ involves a sum over a number of independent indices equal to the number of cycles of $\pi$, each yielding a factor of $q$.

From this interpretation of the inner product \eqref{eq:IPdef}, we are required to take $q$ to be an integer since it is the dimension of an auxiliary Hilbert space. Given this restriction, we are also guaranteed that $\eta_{p,q}^{(n)}$ is positive semi-definite from the general comments above. Positivity is no longer guaranteed if we relax this to allow non-integral $q$, and in fact it can fail (see section \ref{sec:disc}).

Finally, we should move beyond our abstract comments and discuss how such a gauging can arise in gravity. A simple example of this is the `Polchinski-Strominger model' \cite{Polchinski:1994zs,Marolf:2020rpm,Maxfield:2022sio} , in which $\Sym(n)$ appears as a subgroup of diffeomorphisms. For this model we consider $n$ black holes formed from collapse, which have partially evaporated. In perturbation theory, we describe the interior state of each with some $p$-dimensional Hilbert space $\CC^p$; this includes the state of matter which fell in to form the black hole as well as the state of interior Hawking partners created as it evaporates. To compute the inner product on these $n$ black holes, we first allow them to evaporate completely, and assume that no remnant of them remains in the ambient spacetime. In that case, the black hole interiors split off and behave like closed universes (albeit with a small region close to evaporation that cannot be described semi-classically). We are therefore left with an inner product over $n$ sets of Hawking radiation and $n$ closed universes. But now there are diffeomorphisms of space which permute these $n$ closed universes: the usual rules of general relativity would instruct us to gauge these diffeomorphisms (by a non-perturbative extension of the symmetry which imposes the momentum constraints in perturbation theory). Equivalently, we treat the post-evaporation black hole interiors as indistinguishable, with Bosonic statistics. The upshot is that we trace over the radiation in a conventional way, but sum over permutations of the states of interest as well as the late interior Hawking partners produced in the subsequent evaporation. The result is
\begin{equation}
	\llangle j_1,\ldots,j_n|i_1,\ldots,i_n\rrangle = \sum_{\pi\in\Sym(n)} \langle j_1,\ldots,j_n|R(\pi)|i_1,\ldots,i_n\rangle \widetilde{\Tr}\left(\tilde{\rho}^{\otimes n}\,\tilde{R}(\pi)\right),
\end{equation}
where $\tilde{\rho}$ is the reduced density matrix of the late Hawking partners in each black hole, and $R$ and $\tilde{R}$ are the above representations which act by permuting factors. Taking a simple model of Hawking radiation which is maximally entangled on a $q$-dimensional Hilbert space, we recover precisely \eqref{eq:IPdef}. The additional degrees of freedom $\tilde{\hilb}_\mathrm{kin}$ in this model are the late Hawking interior partners; the suppression factors $q^{-|\pi|}$ arise from their entanglement with the exterior Hawking radiation.

As argued in \cite{Marolf:2020rpm,Maxfield:2022sio}, each permutation in this model comes from a simplified version of a replica wormhole, in which the `island' consists of the entire post-evaporation black hole interior. In particular, it leads to a Page curve for the entropy in which $\log q$ is interpreted as the Bekenstein-Hawking entropy of each black hole (though this model is not correct in quantitative detail).

We would like to give a similar interpretation to the model described in section \ref{ssec:topmot}, and for replica wormholes more generally. We can do this, but it requires some novelties to interpret the gauge symmetry and the extra degrees of freedom $\tilde{\hilb}_\mathrm{kin}$. For replica wormholes, the permutations we sum over do not act on closed baby universes, but instead on a region with a boundary (this boundary is typically a codimension two surface lying on the event horizon) --- an `island'.  We can interpret these permutations as diffeomorphisms of space, but diffeomorphisms with discontinuities (i.e. they act to permute replicas within the island, but act trivially outside the island). So our gauge symmetry interpretation requires a somewhat radical extension of general coordinate invariance to include such discontinuous changes of coordinates. The extra degrees of freedom $\tilde{\hilb}_\mathrm{kin}$ in such  cases can have two parts: matter degrees of freedom on the island which are entangled with degrees of freedom outside (like the interior Hawking partners), and some unknown  modes close to the boundary of the island at distances shorter than the cutoff for our theory of gravity. These modes contribute to the suppression parameter $q$ as a `matter entropy' $S_\mathrm{matter}$ and an `area' $\frac{A}{4G_N}$ respectively, combining into a `generalised entropy' $S_\mathrm{gen}=\frac{A}{4G_N}+S_\mathrm{matter}$, so $q\sim \exp(S_\mathrm{gen})$. See \cite{Maxfield:2022sio} for further discussion of these ideas.

\section{Null states and Young diagrams}\label{sec:null}

Having defined the model of interest, our aim in the rest of the paper is to identify the null states in $\hilb_{p,q}^{(n)}$, and to count them (or equivalently to count the number of remaining physical states). Null states are elements of $(\CC^p)^{\otimes n}$ with zero norm, or in other words the kernel of the matrix of inner products $\eta_{p,q}^{(n)}$. The dimension of the physical Hilbert space is then given by the rank of this matrix, so we would like to calculate
\begin{equation}\label{eq:dnpq}
	d_{p,q}^{(n)} := \dim \hilb_{p,q}^{(n)} = \rank \eta_{p,q}(\lambda).
\end{equation}

 It is straightforward to characterise these null states with the help of a little representation theory. The perturbative Hilbert space $(\CC^p)^{\otimes n}$ is naturally a representation of two different groups. First, the symmetric group $\Sym(n)$ acts by permuting the $n$ factors. Second, the unitary group $U(p)$ acts as the same change of basis in each $\CC^p$ factor. Furthermore, the action of these two groups commutes. This tells us that $(\CC^p)^{\otimes n}$ decomposes as a direct sum of sectors, each of which is a tensor product of irreducible representations of the two groups. The concrete statement of this decomposition is Schur-Weyl duality:
\begin{equation}\label{eq:SW}
	(\CC^p)^{\otimes n} = \bigoplus_{\lambda \partition n} V_\lambda^{\Sym(n)} \otimes V_\lambda^{U(p)} \,.
\end{equation}
Here, $V_\lambda^{\Sym(n)}$ and $V_\lambda^{U(p)}$ are vector spaces carrying irreducible representations of $\Sym(n)$ and $U(p)$ respectively. 
These are both labelled by partitions $\lambda$ of $n$ (denoted $\lambda \partition n$) into at most $p$ parts, or equivalently Young diagrams with $n$ boxes and at most $p$ rows.

More concretely, each Young diagram corresponds to a wavefunction $\psi_{i_1 i_2\cdots i_n}$ in $(\CC^p)^{\otimes n}$ with specified symmetry properties: each box corresponds to one of the $n$ indices, we symmetrise over indices with boxes in the same row, and we antisymmetrise over indices with boxes in the same column. In particular, the restriction to at most $p$ rows appears because no nonzero wavefunction can be antisymmetric over more indices than the number of values each index can take. $\Sym(n)$ acts to permute wavefunctions with the same pattern of (anti)symmetrisation, but with different sets of indices. We will not need to know much more about Schur-Weyl duality except for the dimensions of the irreducible representations (given in the next section); more details are doubtless contained in your favourite representation theory textbook \cite{Fuchs:1997jv}.

Using this description, our first main result gives a simple characterisation of the null states: they are precisely those which fall into representations corresponding to Young diagrams with more than $q$ rows. 
\begin{equation}\label{eq:nullstates}
	\begin{aligned}
		\text{Null states $\ker\eta_{p,q}^{(n)}$:} \qquad &\lambda \partition n \text{ with more than $q$ rows.}\\
		\text{Physical states $\hilb_{p,q}^{(n)}$:} \qquad &\lambda \partition n \text{ with at most $q$ rows.}
	\end{aligned}
\end{equation}
Null states correspond simply to wavefunctions which are antisymmetric over too many indices.

We can explain this result from the description of the inner product given in \ref{ssec:gaugemot}, as symmetrising over an extended Hilbert space $(\CC^p)^{\otimes n}\otimes (\CC^q)^{\otimes n}$ before tracing out the second factor. The symmetrisation projects the extended Hilbert space onto its singlet sector under $\Sym(n)$. This projection requires any irreducible representation of $\Sym(n)$ in $(\CC^p)^{\otimes n}$ to be paired with the same representation in $(\CC^q)^{\otimes n}$: the trivial representation only appears in the tensor product of two irreps when they are dual, and all irreps of $\Sym(n)$ are real (equivalent to their duals). This means that a Young diagram $\lambda \partition n$ which is absent in the decomposition of $(\CC^q)^{\otimes n}$ is projected out in $(\CC^p)^{\otimes n}$ by $\eta_{p,q}^{(n)}$, and becomes a null state. Hence, since the Scur-Weyl decomposition of $(\CC^q)^{\otimes n}$ contains only $\lambda$ with at most $q$ rows, this restriction is inherited by the physical Hilbert space $\hilb_{p,q}^{(n)}$.

We explain more details of this result in appendix \ref{app:reptheoryIP}. Additionally, using the grand orthogonality theorem and the expression \eqref{eq:Gaverage2} for the inner product we find an explicit expression for  the matrix of inner products $\eta_{p,q}^{(n)}$ using the Schur-Weyl decomposition \eqref{eq:SW}. Since $\eta_{p,q}^{(n)}$ commutes with the action of both $U(n)$ and $\Sym(n)$, Schur's lemma tells us that it acts as a constant $\eta_{p,q}^{(n)}(\lambda)$ in each sector of fixed $\lambda$; the constant is given in terms of dimensions of representations by
\begin{equation}\label{eq:etalambda}
	\eta_{p,q}^{(n)}(\lambda) = \frac{n!}{q^n}  \frac{\dim V_\lambda^{U(q)}}{\dim V_\lambda^{\Sym(n)}}.
\end{equation}
In particular, this constant is zero precisely when $\lambda$ has more than $q$ rows.

With this characterisation of the null states, we have a formula for the dimension \eqref{eq:dnpq} of the physical Hilbert space $\hilb_{p,q}^{(n)}$:
\begin{equation}\label{eq:dimreps}
\begin{gathered}
	d_{p,q}^{(n)} = \sum_{\substack {\lambda\vdash n \\ \#\text{rows}\leq q}} \dim V^{\Sym(n)}_\lambda \dim V^{U(p)}_\lambda .
\end{gathered}
\end{equation}
To understand how this quantity behaves, we need expressions for the dimensions of the irreps appearing, which we turn to now.

\section{Dimensions from a Coulomb gas}\label{sec:Coulomb}

We have seen that counting physical states amounts to counting states which fall into certain irreducible representations of $U(p)$ and $\Sym(n)$. Fortunately, the dimensions of these representations have expressions which are rather intuitive for a physicist; from this we will find an emergent collective description of the Young diagrams which governs our state counting problem in the most interesting regime of large $n,p,q$.

The main ideas we use originated \cite{vershik1985asymptotic,logan1977variational} in studies of a mathematical problem of random permutations (described in \ref{sec:simplify} and appendix \ref{app:LIS}). A problem more closely related to our state counting was studied in \cite{biane2001approximate}, though our precise problem of interest seems to be new.\footnote{The cited paper determined the typical shape of a Young diagram for a random state in $(\CC^p)^{\otimes n}$, but not with the constraint on the number of rows.} The exact formulation of these ideas that we use follows Okounkov \cite{okounkov2001infinite}.

\subsection{Dimensions of irreducible representations}

We first introduce a convenient set of variables (the `modified Frobenius coordinates') to identify a Young diagram $\lambda$ with at most $q$ rows. This parameterises each diagram with an increasing sequence of $q$ integers $\lambda_i$, defined by
\begin{equation}\label{eq:lambdai}
	\lambda_i = i-(\text{length of $i$th row of $\lambda$}) \quad\implies \quad \lambda_1<\lambda_2<\cdots<\lambda_q\leq q.
\end{equation}
This is increasing since each row is no longer than the row above it.
If $\lambda$ has $r<q$ rows, we simply take the length of an absent row to be zero, so $\lambda_i=i$ for $i>r$. The total number of boxes is given in terms of the sum over all $\lambda$, as
\begin{equation}\label{eq:sumlambda}
	n = \tfrac{1}{2}q(q+1)-\sum_{i=1}^q\lambda_i \,.
\end{equation}

\begin{figure}
\centering
	\includegraphics[width=.7\textwidth]{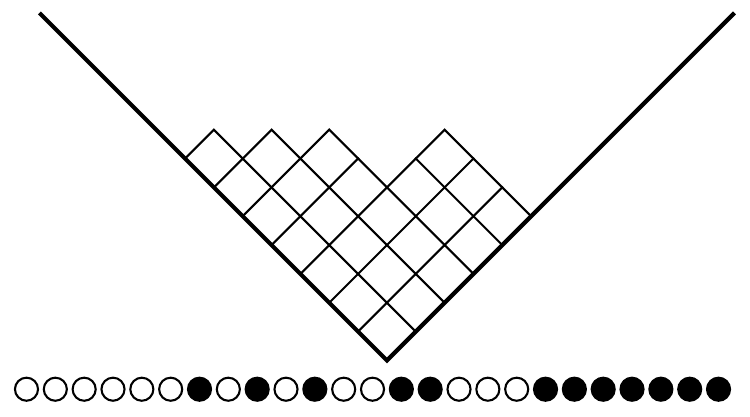}
	\captionsetup{singlelinecheck=off}
    \caption[.]{
    An illustration of the $\lambda_i$ coordinates for a Young diagram introduced in \eqref{eq:lambdai}. We look at the example $\lambda = (7,6,5,3,3) \partition 24$:
    \begin{displaymath}
      \lambda = \raisebox{11pt}{\ydiagram{7,6,5,3,3}}\quad \longrightarrow \lambda_i= \{-6,-4,-2,1,2,6,7,\ldots,q\}.
    \end{displaymath}
    These $\lambda_i$ can be interpreted by rotating the diagram 135\textdegree\  from the usual `English' presentation to the `Russian' presentation in the figure, so the edge of the diagram is a sequence of edges $\diagup$ (denoted by `particles' $\bullet$) and $\diagdown$ (denoted by `holes' $\circ$). The $\lambda_i$ are the locations of the particles in this sequence. The dimensions of representation of $\Sym(n)$ and $U(p)$ corresponding to $\lambda$ are determined by a Coulomb gas Hamiltonian $\mathsf{H}$ for these particles.\label{fig:YoungFermion}
    }
\end{figure}

The set of $\lambda_i$ has a nice interpretation in terms of the geometry of the diagram $\lambda$, which is revealed by rotating the diagram by 135\textdegree as shown in figure \ref{fig:YoungFermion}. The boundary of the Young diagram is then given by a sequence of diagonal lines, with $\diagup$ at the end of a row and $\diagdown$ at the bottom of a column. The $\lambda_i$ simply give the positions of the $\diagup$ segments in the sequence. We can also notate this as a sequence of filled circles $\bullet$ appearing at positions $\lambda_i$ and empty circles $\circ$  at all other positions $i\leq q$. It will be helpful to think of these as `particles' $\bullet$ and `holes' (or the absence of a particle) $\circ$ which lie on a discrete lattice labelled by integers $i\leq q$. We then have a total of $q$ particles which we can distribute over the lattice, and their `dipole moment' $\sum_i\lambda_i$ determines the number of boxes $n$.
\ytableausetup{boxsize=6pt}

The particle interpretation comes into its own once we give expressions for the dimensions of representations in these variables. First the dimension of a representation of the symmetric group is given by
\begin{equation}\label{eq:dimSym}
	\dim V^{\Sym(n)}_\lambda = \frac{n!}{\prod_{i=1}^q {(q-\lambda_i)!}} \prod_{1\leq i<j\leq q} (\lambda_j-\lambda_i).
\end{equation}
The final factor $\prod (\lambda_j-\lambda_i)$ is the Vandermonde determinant of the $\lambda_i$. As may be familiar from random matrix theory, this has a nice interpretation as the interaction potential of a Coulomb gas. Specifically, we can write it as the exponential of minus $-\sum\log(\lambda_j-\lambda_i)$ (summed over pairs), which is the interaction energy of particles located at positions $\lambda_i$ with a repulsive force proportional to the inverse of their separation. Similarly, we can interpret the factor $\prod_{i=1}^q {(q-\lambda_i)!}$ as providing a potential $V(\lambda) = \log(q-\lambda)!$ for each particle.

We have a similar expression for the dimension of a representation of $U(p)$,
\begin{equation}\label{eq:dimU}
	\dim V^{U(p)}_\lambda = \frac{1}{\prod_{i=1}^q {(p-i)!}}  \prod_{i=1}^q \frac{(p-\lambda_i)!}{(q-\lambda_i)!} \prod_{1\leq i<j\leq q} (\lambda_j-\lambda_i),
\end{equation}
where we have assumed $p\geq q$.\footnote{This formula for dimensions of $U(p)$ representations has a particularly simple interpretation in terms of the particle energy which we can see by taking $p=q$. Then, the $\lambda$ dependence comes from the Vandermonde, so the dimension is computed by only the Coulomb interaction energy of $p$ particles located on the lattice of integers $i\leq p$ (with no external potential). This is measured relative to the $\lambda$-independent prefactor, which simply subtracts the energy of the maximally energetic configuration  (all particles maximally to the right, $\lambda_i=i$) corresponding to the empty Young diagram (i.e., the trivial dimension one representation). The extra factor for $q<p$ accounts for the energy due to the $q-p$ constrained particles with $\lambda_i=i$ ($q+1\leq i\leq p$). In particular, since the dimension depends only on the separation of the $p$ particles, it is invariant under translation: translating all particles to the left adds an extra column of height $p$ to the diagram, which corresponds to taking the tensor product with the one-dimensional determinant representation of $U(p)$.}

Combining these, we can write the dimension of our Hilbert space in the form of a classical statistical partition function for a Coulomb gas of $q$ particles,
\begin{equation}\label{eq:partitionfunction}
	d_{p,q}^{(n)} = \sum_{\substack {\lambda_1 <\cdots <\lambda_q\leq q \\ \lambda\partition n}} e^{-\mathsf{H}_{p,q}^n(\lambda)}
\end{equation}
with `Hamiltonian'
\begin{equation}\label{eq:dimHam}
	\mathsf{H}_{p,q}^n(\lambda)= -2\sum_{1\leq i<j \leq q}\log(\lambda_j-\lambda_i) +\sum_{i=1}^q \mathsf{V}(\lambda_i) + \mathsf{H}_0,
\end{equation}
where the `potential' $\mathsf{V}$ is given by
\begin{equation}
	\mathsf{V}(\lambda) = 2\log(q-\lambda)! - \log(p-\lambda)!
\end{equation}
and the constant shift $\mathsf{H}_0$ is
\begin{equation}
	\mathsf{H}_0 = \sum_{i=1}^q \log(p-i)! - \log n! \,.
\end{equation}
The sum over $\lambda$ is restricted to partitions of $n$, which means that the sum \eqref{eq:partitionfunction} runs only over configurations satisfying the constraint \eqref{eq:sumlambda} on the `dipole' $\sum_i\lambda_i$ of the particles.

We emphasise that the Hamiltonian $\mathsf{H}$ is not related to the physical energy or time evolution in our problem.

\subsection{Continuum limit}

Our interest lies mostly in the limit of large $n,p,q$. This has two effects on the system of particles above. First, instead of discrete particles at integer locations we can approximate them as a continuum with some density $\rho$. Second, the large parameter limit is effectively a low temperature limit for our partition function \ref{eq:partitionfunction}, so the dimension will be dominated by configurations close to the `ground state' that minimises $\mathsf{H}_{p,q}^n(\lambda)$.

The most interesting regime occurs when we take $n\to\infty$ and with $p,q$  of order $\sqrt{n}$. In that case, the typical Young diagram will have $O(\sqrt{n})$ rows and columns, so the $\lambda_i$ will also scale as $\sqrt{n}$. With this in mind, we study a limit governed by a small positive parameter $\epsilon$, holding fixed the quantities
\begin{equation}
	\hat{n} = \epsilon^2 n, \quad \hat{p} = \epsilon p, \quad \hat{q} = \epsilon q
\end{equation}
while taking $\epsilon\to 0$. We then define rescaled $\lambda_i$ by
\begin{equation}
	x_i = \epsilon \lambda_i,
\end{equation}
and encode their values in a `density' defined as
\begin{equation}
	\rho(x) = \epsilon \sum_i \delta(x-x_i).
\end{equation}

In the $\epsilon\to 0$ limit, we can approximate the density $\rho$ describing a typical Young diagram by a continuous function on the region $x<\hat{q}$. Importantly, the fact that $\lambda_i$ are distinct integers means that the density of particles cannot be arbitrarily large, with $\rho$ bounded as
\begin{equation} \label{eq:rhoineq}
	0\leq \rho(x)\leq 1.
\end{equation}
We also have equality constraints on $\rho$ coming from setting the number of $\lambda_i$ to $q$ and the number of boxes to $n$, giving
\begin{align}
	\int \rho(x) dx &= \hat{q}, \label{eq:rhoConstraint} \\
	\int x\rho(x) dx &\sim \tfrac{1}{2}\hat{q}^2-\hat{n}, \label{eq:dipole}
\end{align}
where the second equation is the limiting version of \eqref{eq:sumlambda}.

If we now take $\epsilon\to 0$ while holding fixed $\hat{p}$, $\hat{q}$, $\hat{n}$ and the function $\rho$, our dimension counting Hamiltonian \eqref{eq:dimHam} grows as $\epsilon^{-2}$: 
\begin{equation}
	\mathsf{H}(\lambda) \sim \frac{1}{\epsilon^2}\hat{\mathsf{H}}[\rho]  - n\log p \,.
\end{equation}
We have extracted the shift $n\log p = \epsilon^{-2} \log\left(\hat{p}/\epsilon\right)$ so that we recover the dimension $p^n$ of $(\CC^p)^{\otimes n}$ (with no constraint on the number of rows of $\lambda$) when $\hat{\mathsf{H}}=0$, and otherwise $\hat{\mathsf{H}}$ is positive. The limiting energy functional $\hat{\mathsf{H}}$ is
\begin{equation}\label{eq:dimHam2}
	\hat{\mathsf{H}}[\rho]\sim -\int \log|x_1-x_2| \rho(x_1) \rho(x_2) dx_1 dx_2 + \int \hat{\mathsf{V}}(x) \rho(x) dx + \hat{\mathsf{H}}_0,
\end{equation}
with the potential
\begin{equation}\label{eq:Vhat}
	\hat{\mathsf{V}}(x) =2 (\hat{q}-x) (\log (\hat{q}-x)-1)-(\hat{p}-x) (\log (\hat{p}-x)-1)
\end{equation}
and
\begin{equation}\label{eq:hatH0}
	\hat{\mathsf{H}}_0= \tfrac{1}{2}\hat{p}^2\log\hat{p}-\tfrac{1}{2}(\hat{p}-\hat{q})^2\log(\hat{p}-\hat{q})-\tfrac{3}{4}\hat{q}(2\hat{p}-\hat{q}) -\hat{n} \log\tfrac{\hat{n}}{\hat{p}} + \hat{n} \,.
\end{equation}
The first term in \eqref{eq:dimHam2} (quadratic in $\rho$) comes from the double sum Coulomb interaction term\footnote{Some care is required for the interaction term: the integral $\epsilon^{-2}\int \log|x_1-x_2| \rho(x_1) \rho(x_2) dx_1 dx_2$ differs from the sum $2\sum\log(\lambda_j-\lambda_i)$ by roughly $q^2\log\epsilon$.}, while the potential $\hat{\mathsf{V}}$ is a limit of $\mathsf{V}$, with constant terms and  linear terms in $x$ absorbed into $\hat{\mathsf{H}}_0$ by making use of the constraints \eqref{eq:rhoConstraint} and \eqref{eq:dipole}.

In the partition function \eqref{eq:partitionfunction}, the scaling of $\mathsf{H}$ in this limit means that $\epsilon^2$ plays the role of a temperature. The $\epsilon\to 0$ limit is therefore a low-temperature limit in which the partition function is dominated by the states of small energy. To count states, our task is therefore to compute the minimum of $\hat{\mathsf{H}}$ subject to the constraints. These constraints can be imposed by adding Lagrange multipliers (or chemical potentials); these add terms to the potential which are constant and linear in $x$. Explicitly, we minimise
\begin{equation}
	\hat{\mathsf{H}}[\rho] +\mu \int x \rho(x)dx +\nu\int \rho(x) dx
\end{equation}
subject only to the inequality constraints $0\leq \rho(x) \leq 1$, and subsequently can determine $\mu$, $\nu$ in terms of $\hat{n},\hat{q}$. The chemical potentials $\mu,\nu$ also tell us how the minimum energy $ \hat{\mathsf{H}}_\mathrm{min}$ depends on parameters: in particular we have
\begin{equation}\label{eq:dHdn}
	\frac{\partial \hat{\mathsf{H}}_\mathrm{min}}{\partial \hat{n}} = \mu - \log\left(\frac{\hat n}{\hat{p}}\right),
\end{equation}
 where the shift comes from the explicit  $n$-dependence of $\hat{\mathsf{H}}[\rho]$ through $\hat{\mathsf{H}}_0$ in \eqref{eq:hatH0}.

A nice physical picture for $\hat{\mathsf{V}}$ is a regulated potential from particles with maximal density $\rho=1$ in the region $\hat{q}<x<\hat{p}$, and half-maximal density $\rho=\frac{1}{2}$ in the region $x>\hat{p}$: the actual potential arising from these would diverge, but the divergences are constant and linear in $x$ so can be absorbed in the Lagrange multipliers $\mu,\nu$.

\subsection{A simplifying limit}\label{sec:simplify}

For black holes, in the most interesting and typical case there is a large discrepancy between the perturbative entropy and the Bekenstein-Hawking entropy, $S_\mathrm{pert}-S_\mathrm{BH}\gg 1$. This corresponds to the parameter regime $p \gg q$, which affords some simplifications.

If we take $p\to \infty$ for a fixed partition $\lambda$, the dimension for the corresponding representation of $U(p)$ is asymptotic to $\frac{p^n}{n!}$ times the dimension of the corresponding representation of $\Sym(n)$, which can be seen from our formulas \eqref{eq:dimU} and \eqref{eq:dimSym}. From this, the number of states $\dim V_\lambda^{\Sym(n)} \dim V_\lambda^{U(p)}$ falling into such representations approaches $p^n$ times the `Plancherel measure'  $\mu_n(\lambda)$ for $\Sym(n)$,  a probability measure proportional to the square of the dimensions of representations of $\Sym(n)$:
\begin{equation}\label{eq:Planch}
	\dim V_\lambda^{\Sym(n)} \dim V_\lambda^{U(p)} \sim p^n \mu_n(\lambda), \qquad \mu_n(\lambda) = \frac{1}{n!}\left(\dim V_\lambda^{\Sym(n)}\right)^2.
\end{equation}
In this limit, our state counting exercise becomes closely connected to the venerable `longest increasing subsequence' problem in combinatorics, first studied by Ulam. We briefly explain the connection in appendix \ref{app:LIS}.

The Plancherel measure approximation is good for $n\ll p q$. Once $n$ is of order $p q$, the typical length of rows of a Young diagram will become of order $p$, meaning that we will have some (negative) $\lambda_i$ of order $p$. In that regime, we can no longer approximate using the Plancherel measure, though the problem will simplify for different reasons as we will see in the next section.

\section{State counting} \label{sec:counting}

We saw above that most states in our physical Hilbert space $\hilb_{p,q}^{(n)}$ lie in representations described by Young diagrams close to a particular shape. The dominant shape is determined by minimising the Coulomb gas Hamiltonian $\hat{\mathsf{H}}[\rho]$ given in \eqref{eq:dimHam2}. This problem is solved by standard methods \cite{Eynard:2015aea} reviewed in appendix \ref{app:Coulomb}. In this section we describe how the minimising solution $\rho$ and the dimension $d_{p,q}^{(n)}$ behave as we vary the parameters. For ease of presentation we focus on the most interesting case $p\gg q$, with simplifications as outlined in section \ref{sec:simplify}, though full solutions where $p$ and $q$ are of the same order are obtained in appendix \ref{app:Coulomb}. 

\begin{figure}
\centering
\begin{tikzpicture}
		\node at (0,0){\includegraphics[width=.9\textwidth]{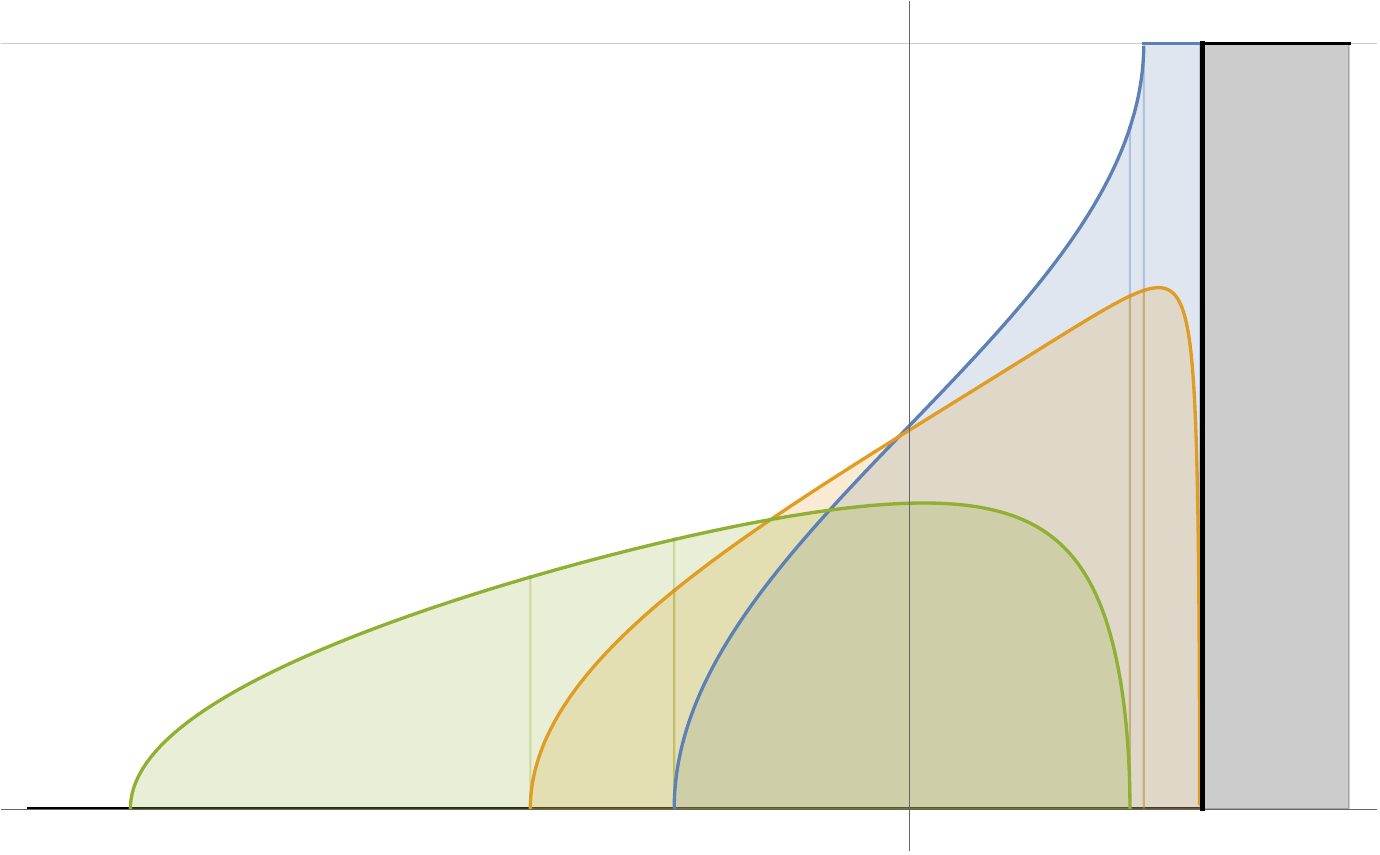}};
		\node at (5.2,-4.2){$x=\hat{q}$};
		\node at (2.2,-4.2){$x=0$};
		\node at (3.6,2.4){(a)};
		\node at (1,.1){(b)};
		\node at (-3.,-1.5){(c)};
	\node at (3.2,2.4){\includegraphics[width=.12\textwidth]{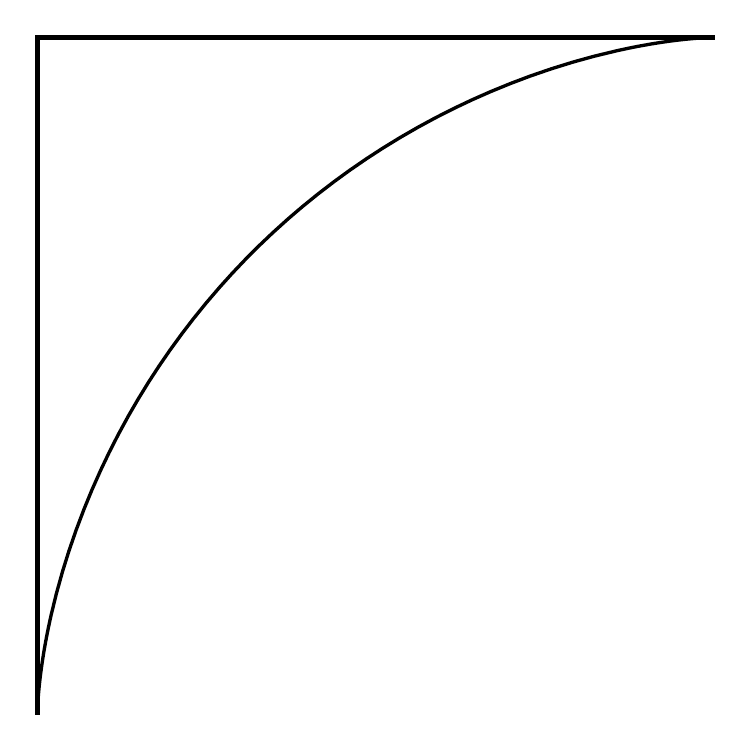}};
		\node at (.8,0.3){\includegraphics[width=.18\textwidth]{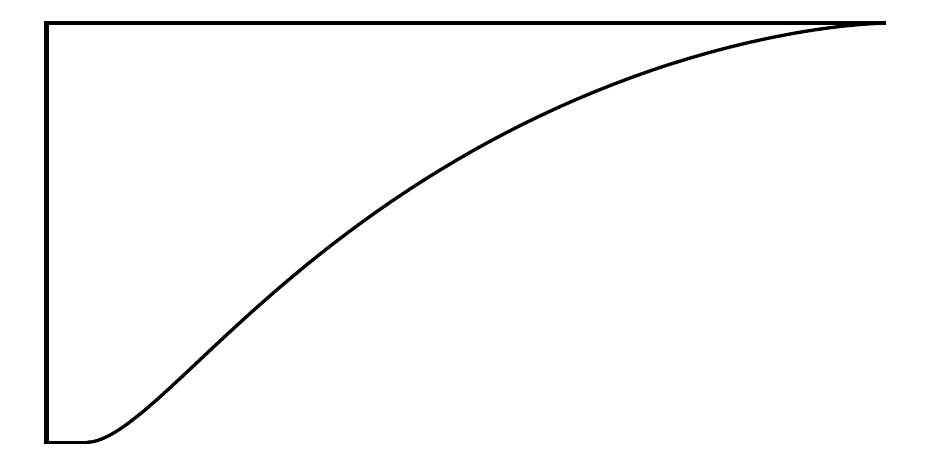}};
		\node at (-3.3,-1.2){\includegraphics[width=.44\textwidth]{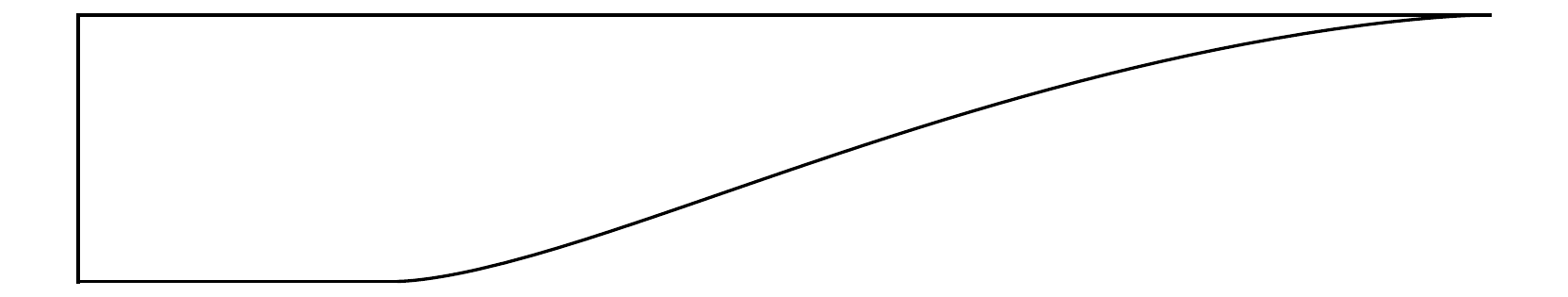}};
\end{tikzpicture}
    \caption{The equilibrium density $\rho(x)$ for various values of $\frac{n}{q^2}$ (taking $p\gg q$), along with the corresponding shapes of Young diagrams. The grey region $x>\hat{q}$ is required to have maximum density $\rho=1$, corresponding to the upper bound $q$ on the number of rows of the Young diagram. In case (a) (in blue) we have the density described in \eqref{eq:arcsin} with $n<\frac{q^2}{4}$: the upper bound $\rho(x)\leq 1$ is saturated in the region $\sqrt{2\hat{n}}<x<\hat{q}$, which means that the corresponding Young diagrams have fewer than $q$ rows and there are exponentially few null states. Increasing $n$, we pass through a phase transition to a density shown in (b) (in yellow). There is a small gap of zero density to the right, which means that the corresponding Young diagram has a flat bottom. As $n$ increases, the density broadens and moves to the left as shown in case (c) (green), so the length of the last row of the Young diagram increases. Ultimately for $n \gg q^2$, the density approaches a semicircle centred at $x\sim -\frac{\hat{n}}{\hat{q}}$.}
    \label{fig:CoulombSolPlots}
\end{figure}
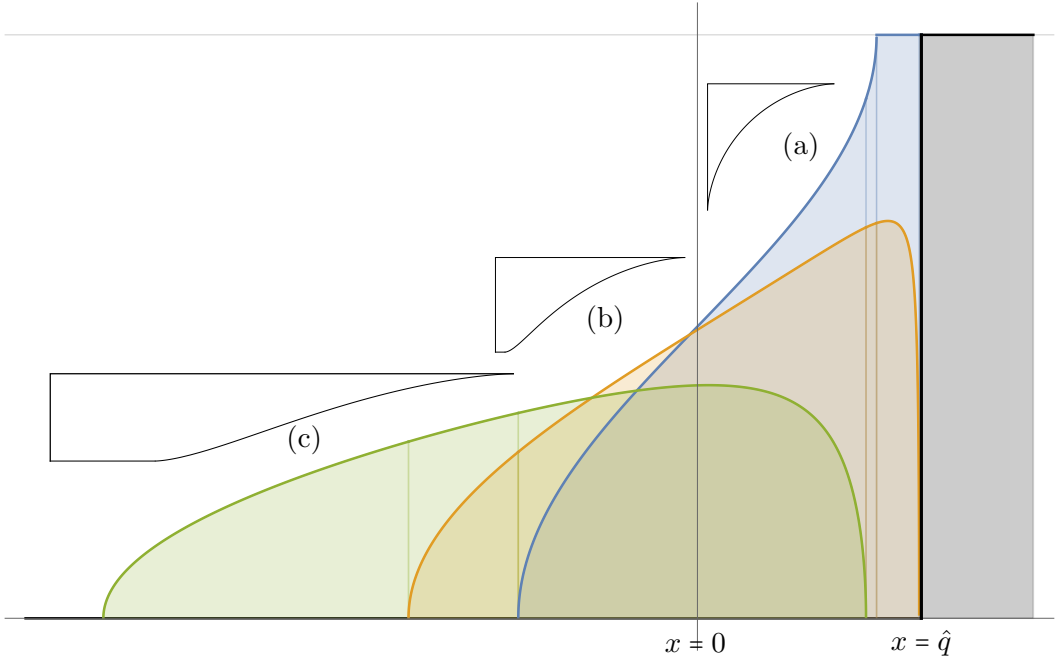

For small $n \lesssim q^2$, the constraint on the number of rows is not saturated for the typical Young diagram: the $\lambda$ which maximises $\dim V^{\Sym(n)}_\lambda \dim V^{U(p)}_\lambda$ has $r$ rows for some $r<q$. While there are some null states for $n>q$, their number is exponentially small (relative to the total dimension $p^n$) until $n$ is of order $q^2$. In terms of the density $\rho$, this means that its upper bound is saturated with $\rho(x)=1$ for an interval $\hat{r} < x< \hat{q}$, where $\hat{r} = \epsilon r$. Specifically, for $n \ll p^2$ and $n \leq \frac{q^2}{4}$ we have $\hat{r}\sim 2\sqrt{\hat{n}}$, and $\rho$ is given by a simple inverse sine:
\begin{equation}\label{eq:arcsin}
	\rho(x) = \begin{cases}
		0 & x \leq  - 2\sqrt{\hat{n}} \\
		\frac{1}{2} + \frac{1}{\pi}\sin^{-1}\left(\frac{x}{2\sqrt{\hat{n}}}\right) & - 2\sqrt{\hat{n}} < x<2\sqrt{\hat{n}} \\
		1 & x \geq 2\sqrt{\hat{n}}
	\end{cases}
\end{equation}
This result is famous as the shape of the typical Young diagram chosen from the Plancherel measure \eqref{eq:Planch} \cite{logan1977variational,vershik1985asymptotic}. This density (along with a sketch of the corresponding Young diagram) is shown as the blue curve in figure \ref{fig:CoulombSolPlots}.

As we increase $n$, the region where $\rho$ lies strictly between $0$ and $1$ broadens, and eventually the bound on the number of rows will become relevant. At this point there is a phase transition in the density $\rho$. For $p\gg q$ this occurs when $n =\frac{q^2}{4}$, after which the inverse sine solution \eqref{eq:arcsin} is no longer admissible. For $n>\frac{q^2}{4}$ (but still $n \ll p q$ so that we can use the Plancherel approximation of section \ref{sec:simplify}), the density $\rho$ is supported on an interval $[x_-,x_+]$, with 
\begin{equation}
	x_\pm = \frac{\hat{q}}{4}-\frac{\hat{n}}{\hat{q}} \pm \sqrt{2 \hat{n}+\tfrac{1}{2}\hat{q}^2}.
\end{equation}
The energy-minimising density in this interval becomes
\begin{equation}\label{eq:rhoResult}
	\rho(x) = \frac{1}{\pi}\sin^{-1} f_\infty(x)-\frac{1}{\pi}\sin^{-1}f_{\hat{q}}(x),
\end{equation}
where $f_X$ is the fractional linear (M\"obius) map for which $f_X(x_\pm)=\pm 1$ and $f_X(X)=\infty$ (an explicit expression for $f_X$ is given in \eqref{eq:fX}). Note that if we take $\hat{n} \to \frac{\hat{q}^2}{4}$, this becomes $x_\pm=\hat{q}$, and the density $\rho(x)$ approaches \eqref{eq:arcsin}. However, the qualitative behaviour at the right edge $x\to x_+$ changes: for $\hat{n}>\frac{\hat{q}^2}{4}$, $\rho$ vanishes with a square-root edge $\rho(x)\propto \sqrt{x_+-x}$ as $x\to x_+$; for $\hat{n}\leq \frac{\hat{q}^2}{4}$, $\rho$ instead has a similar square-root approach to its upper bound, with $\rho\to 1$ as $x\to 2\sqrt{\hat{n}}$. Plots of this density for various values of $\frac{n}{q^2}$ are shown in yellow and green in figure \ref{fig:CoulombSolPlots}.

By computing the chemical potential $\mu$ for $\hat{n}$ (which imposes the constraint on dipole moment $\int x\rho(x)dx$), we can determine how the minimal energy $\hat{\mathsf{H}}_\mathrm{min}$ of our Coulomb gas depends on $\hat{n}$ in this regime using \eqref{eq:dHdn}, finding
\begin{equation} \label{eq:dHdn1}
	\frac{\partial{\hat{\mathsf{H}}_\mathrm{min}}}{\partial{\hat{n}}} = 2\log \left(\frac{4\hat{n}+\hat{q}^2}{4\hat{q}\sqrt{\hat{n}}}\right).
\end{equation}
This vanishes quadratically at the phase transition $\hat{n} = \frac{\hat{q}^2}{4}$, where
\begin{equation}\label{eq:Hmintrans}
	 \hat{\mathsf{H}}_\mathrm{min} \sim \frac{4}{3\hat{q}^4}\left(\hat{n} - \tfrac{\hat{q}^2}{4}\right)^3, \qquad (0<\hat{n}- \tfrac{\hat{q}^2}{4}\ll \hat{q}^2).
\end{equation}
For $\hat{n}<\frac{\hat{q}^2}{4}$ we have $\hat{\mathsf{H}}_\mathrm{min}=0$, so this cubic dependence indicates a third-order phase transition. This is similar to two famous third-order phase transitions in two-dimensional large-$N$ QCD, the Gross-Witten-Wadia \cite{Gross:1980he,Wadia:1980cp} and Douglas-Kazakov \cite{Douglas:1993iia} transitions, but differs  qualitatively  from both.\footnote{The GWW transition concerns the density of eigenvalues of a unitary matrix, which live on a circle; the density transitions from support on an interval to support on the whole circle. The DK transition concerns Young diagrams corresponding to representations of a $U(N)$ gauge group, arriving at a Coulomb gas in a quadratic potential with an upper bound $\rho\leq 1$ on the density; the transition is between a semicircle and a configuration which saturates the upper bound on an interval. Both transitions involve the merging of two square-root edges in $\rho$, while our transition involves a single such edge moving away from the `boundary' $x=\hat{q}$.}

By integrating \eqref{eq:dHdn1} we can obtain a formula for the dimension $d_{p,q}^{(n)}$ valid in the regime $1\ll \frac{q^2}{4}\leq n\ll p q$. In particular, for $n\gg q^2$ we have a simple behaviour which we write in terms of the `entropy per black hole',
\begin{equation}\label{eq:dimintermediate}
	\frac{1}{n}\log d_{p,q}^{(n)} \sim \log\left(\frac{pq^2}{n}\right) \qquad (q^2\ll n\ll pq),
\end{equation}
up to an additive correction of order unity. This is much smaller than the perturbative entropy $S_\mathrm{pert} = \log p$, but still much larger than the Bekenstein-Hawking entropy $S_\mathrm{BH}=\log q$ while $n\ll p q$.

To see what happens when $n$ is of order $pq$ or larger, we need to use a different approximation. Fortunately, this is rather simple and intuitive. For large $\hat{n}$, we require that the dipole $\int x\rho(x)dx \sim -\hat{n}$ becomes large and negative, with fixed total density $\int \rho = \hat{q}$. This requires our Coulomb gas blob to move leftwards, centred around $x \sim -\frac{\hat{n}}{\hat{q}}$.  To achieve this, we can simply tune our chemical potential $\mu$ so that $\hat{\mathsf{V}}(x)+ \mu x$ has a minimum at that value of $x$, meaning that we choose $\mu\sim -\hat{\mathsf{V}}'(-\frac{\hat{n}}{\hat{q}})$. Then, if $\hat{\mathsf{V}}''(x)$ is sufficiently large the density will be confined close to that minimum, with the width $x_+-x_-$ much smaller than $\frac{\hat{n}}{\hat{q}}$.

It turns out that this is a good approximation for $n\gg q^2$ (any scaling with $p$ allowed). Furthermore, in that regime we can use a quadratic approximation for the potential $\hat{\mathsf{V}}(x)+ \mu x$ near its minimum. As a result, the density $\rho(x)$ approaches the famous Wigner semicircle form, with width $x_+-x_-= 4\sqrt{\frac{\hat{q}}{\hat{\mathsf{V}}''}}$. And indeed, the expression \eqref{eq:rhoResult} for $\rho(x)$  obtained above approaches a semicircle in the limit $n\gg q^2$. In this regime $\hat{\mathsf{V}}''(-\frac{\hat{n}}{\hat{q}})$ is always of order $\frac{\hat{q}}{\hat{n}}$, so the width  is of order $\sqrt{\hat{n}}$ (in particular, much smaller than $\frac{\hat{n}}{\hat{q}}$ as required).

Using the formula $\mu\sim -\hat{\mathsf{V}}'(-\frac{\hat{n}}{\hat{q}})$ which fixes the minimum of the potential and \eqref{eq:dHdn}, we can immediately read off the $\hat{n}$ dependence of the minimum energy:
\begin{equation} \label{eq:dHdn2}
	\frac{\partial{\hat{\mathsf{H}}_\mathrm{min}}}{\partial{\hat{n}}} \sim \log \left(\frac{\hat{p}\hat{n}}{\hat{q}(\hat{n}+\hat{p}\hat{q})}\right).
\end{equation}
For $q^2 \ll n \ll pq$ this agrees with \eqref{eq:dHdn2}. But more interesting is the crossover to the regime $\hat{n}\gg \hat{p}\hat{q}$, which gives us our final large-$n$ limit of the entropy per black hole (a formula which applies not only for $p\gg q$, but whenever $p>q$):
\begin{equation}\label{eq:dimFinal}
	\frac{1}{n}\log d_{p,q}^{(n)} \sim \log q \qquad ( n\gg pq, q<p).
\end{equation}
This means that the number of independent physical states is indeed governed for large $n$ by the Bekenstein-Hawking entropy $S_\mathrm{BH} = \log q$. 

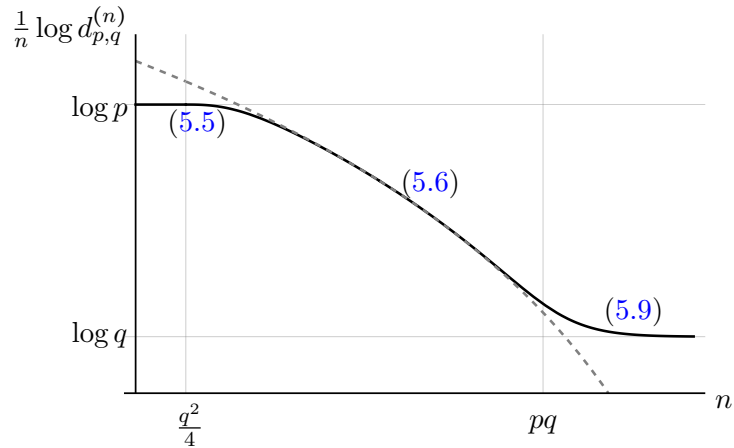
\begin{figure}
\centering
\begin{tikzpicture}
	\node at (-4.6,2.5){$\frac{1}{n}\log d_{p,q}^{(n)}$};
	\node at (4.1,-2.5){$n$};
	\node at (-4.2,1.4){$\log p$};
	\node at (-4.2,-1.6){$\log q$};
	\node at (-3.0,-2.8){$\tfrac{q^2}{4}$};
	\node at (1.7,-2.8){$pq$};
	\node at (0.2,0.4){\eqref{eq:dimintermediate}};
	\node at (-2.9,1.2){\eqref{eq:Hmintrans}};
	\node at (2.9,-1.3){\eqref{eq:dimcorrections}};
	\node at (0,0){\includegraphics[width=.5\textwidth]{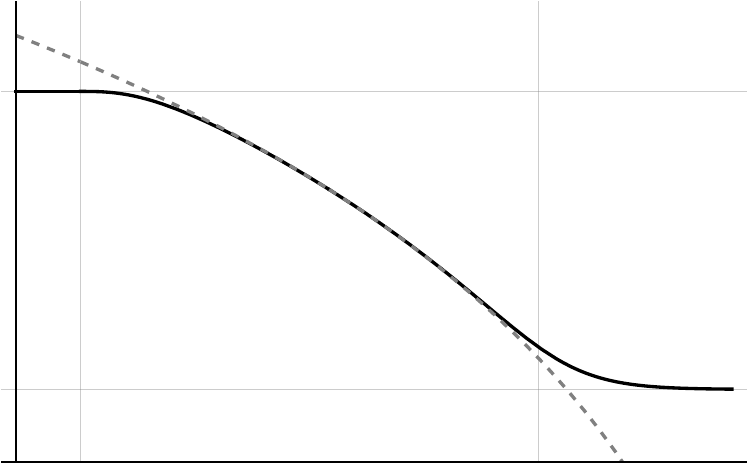}};
\end{tikzpicture}
    \caption{
    The dimension of the physical Hilbert space in the form of the `entropy per black hole' $\frac{1}{n}\log d_{p,q}^{(n)}$, plotted as a function of $n$. We have used logarithmic scales on both axes to show behaviour as quantities change over many orders of magnitude. Various regimes are indicated with equation numbers, pointing to the relevant discussion in the text. The dashed curve is $\log\left(\frac{p q^2}{n}\right)$ as in \eqref{eq:dimintermediate}, governing the intermediate regime $q^2\ll n \ll p q$.
    \label{fig:dimensionPlot}
    }
\end{figure}

By going to the next order in the large $n$ expansion, we can see how quickly this asymptotic dimension is approached. The corrections are given by
\begin{equation}\label{eq:dimcorrections}
	\frac{1}{n}\log d_{p,q}^{(n)} \sim \log q -  p q \frac{\log n}{n} + \cdots.
\end{equation}
The heuristic described below \eqref{eq:interpent} is in this sense misleading, since the corrections in a theory with a finite number of $\alpha$-sectors would be of order $\frac{1}{n}$, as opposed to the slightly larger $\frac{\log n}{n}$ seen here. We do not have a particularly satsisfying interpretation for this correction. Perhaps the additional $\log n$ could be related to the width of the $\alpha$-state wavefunction as the inner product of EOW brane states is `learned' to higher accuracy with more copies of the system.

The dependence of the dimension on $n$ is plotted in figure \ref{fig:dimensionPlot}.


\section{Discussion}\label{sec:disc}

\subsection{Eigenvalues of the inner product}

We have focussed on calculating the dimension of the physical Hilbert space, or the rank of the inner product $\eta$. But this is not the only relevant information about the inner product: we can also ask whether the norm of physical (non-null) states is dramatically altered relative to the perturbative inner product, by looking at the distribution of non-zero eigenvalues of $\eta$. (Since we can always an overall factor is not physically relevant, we should ask about the relative size of different eigenvalues.) We might hope for the simplest possible answer, that $\eta$ is roughly proportional to the identity matrix on its image (so the inner product is roughly a projection onto the physical states). In fact this does not happen: instead there is a small number of states with very large eigenvalues relative to the typical size.

To study this, we look at the value $\eta_{p,q}^{(n)}(\lambda)$ of the inner product acting on states in the representation $\lambda$, given in \eqref{eq:etalambda} (derived in appendix \ref{app:reptheoryIP}). Using expressions \eqref{eq:dimSym}, \eqref{eq:dimU} for the dimensions of representations, we can write this as
\begin{equation}\label{eq:etanpq}
	\eta_{p,q}^{(n)}(\lambda) = \frac{1}{q^n}   \frac{\prod_{i=1}^q {(q-\lambda_i)!}}{\prod_{i=1}^q {(q-i)!}}.
\end{equation}
In terms of the Coulomb gas, we can think of this as an observable which is the exponential of a function linear in the density $\rho(x)$. It is quite sensitive to changes in the representation $\lambda$: for example, even moving a single `particle' a few sites will change $\eta_{p,q}^{(n)}(\lambda)$ by a factor of $q$ to some power.

As an example to get some handle on this quantity, consider computing $\Tr(\eta^k)$. This is a Coulomb gas $k$-point function, obtained by inserting $(\eta_{p,q}^{(n)}(\lambda))^k$ in the `partition function' \eqref{eq:partitionfunction} before summing over $\lambda$ weighted by the `Boltzmann factors' coming from the dimensions of the  corresponding representations. If the eigenvalues had a narrow distribution in which very large eigenvalues are rare, this would behave as $\dim (\hilb_\mathrm{phys})^{1-k}(\Tr\eta)^k$. In terms of the Coulomb gas, the correlation functions would be dominated by the value in the ground state. This means that $\eta$ is close to a multiple of a projector in an appropriate norm (e.g., Frobienius norm for $k=2$). But this is not the case: instead, these insertions `backreact', altering which $\lambda$ dominate the sum.

We can see this because the insertions take the same form as the dimensions appearing in the original `Hamiltonian'. Specifically, $\eta^k$ adds an additional potential term to  \eqref{eq:Vhat},  $\delta \hat{\mathsf{V}}=-k(\hat{q}-x)(\log(\hat{q}-x)-1)$. This altered potential changes the energy-minimising $\lambda$. For example, if we look at the regime of $n \gg pq$ \eqref{eq:dimFinal}, the second derivative of the modified potential is $(1-k)\frac{\hat{q}}{\hat{n}}$. This gives a stable minimum only for $k<1$! In this regime, the calculation is dominated by a semicircle of width $4 \sqrt{\frac{\hat{n}}{1-k}}$. For $k\geq 1$, the continuum limit is no longer applicable and $(\Tr\eta)^k$ appears to be dominated by Young diagrams with one very long row, having very few states but very large eigenvalue $\eta_{p,q}^{(n)}(\lambda)$.

It would be interesting to understand if this property of $\eta$ has any important consequences. For asymptotic observers this certainly does not make an impact, since in that case we can always use the `ensemble' language of a random factorizing inner product. But it may be significant for physics in the black hole interior.

\subsection{Connection to more realistic models of gravity}\label{ssec:realistic}

One might worry that the model studied here is too simple to take lessons which generalise to more realistic theories (e.g., Einstein gravity in four spacetime dimensions). On the contrary, we argue here that our inner product captures a universal feature present in any theory of gravity in which replica wormholes are present.

The connection is sharpest if we work in `fixed area states' \cite{Akers:2018fow,Dong:2018seb}. For these states we choose a codimension two surface $\Sigma$ in a background spacetime, and specify that the area of this surface has very small fluctuations (i.e., the wavefunction is sharply peaked in directions corresponding to changes in area). In the discussion of JT gravity in section \ref{ssec:topmot}, this was achieved by specifying a narrow microcanonical window of energies. If we compute a quantity involving $n$ copies of such a state, replica wormholes formed by an $n$-fold branched cover of the original spacetime give a stationary action configuration. The branching occurs at $\Sigma$, and is labelled by a permutation $\pi$ of the $n$ replicas. The resulting spacetime has conical singularities at $\Sigma$, but still solves the equations of motion due to the constraint on the area. The classical action is suppressed by a factor $\exp(-|\pi|\frac{\mathrm{Area}(\Sigma)}{4G_N})$, and the branching means that quantum fields on the interior of $\Sigma$ are permuted by $\pi$. Thus, summing over $\pi$ results in precisely the structure of our inner product \eqref{eq:IPdef}.

 By using fixed are states we are not allowing quantum fluctuations of the area: a general state is a superposition of fixed area states of different areas. Fortunately, we can incorporate this in a simple way using the expression for the  inner product  \eqref{eq:etaProj} as a partial trace over auxiliary degrees of freedom. We concentrated on the case in which the state $\tilde{\rho}$ in this auxiliary Hilbert space is maximally mixed, a choice which corresponds to a fixed area state. Fluctuations in area simply correspond to allowing a more general state. This parallels the idea that fixed area states correspond to a `flat entanglement spectrum' \cite{Dong:2018seb}. Ultimately, this does not make any difference to the dimension of the physical Hilbert space. For this, it matters only which representations are present in the support of $\tilde{\rho}$, which is not sensitive to the details of the state.
 
 As a concrete example, in JT gravity we can consider any window of energies as long as it contains a finite number $q$ of states. We can take the auxiliary space $\tilde{\hilb}_\mathrm{kin}$ to be spanned by ($n$ copies of) these $q$ energy eigenstates. The density matrix $\tilde{\rho}$ is then a function of these energies given explicitly in appendix \ref{app:JTIP}, which also depends on the `temperature' we take for our EOW brane states. This $\tilde{\rho}$ is maximally mixed only in the approximation that the energies are all very similar. The inner product will become a more complicated sum over permutations $\pi$ weighted by a class function of $\pi$, but its rank is unchanged.  
 
 While our model is thus present in more general theories of gravity, it is not the full answer since the structure of our inner product includes only a simple class of topologies. It would certainly be interesting to understand the effects of more complicated spacetimes, in particular from wormholes connecting different locations in a single spacetime (so that we are not forced to consider replicas).

\subsection{Unoriented and fermionic universes}

Our model for the inner product was obtained from sums over oriented spacetimes. A natural modification allows unoriented spacetimes (gauging orientation-reversing diffeomorphisms as well as orientation preserving). From the boundary point of view, this corresponds to a dual theory with time-reversal symmetry. The main consequence for the inner product is that we lose the distinction between the $i$ and $j$ indices labelling `bra' and `ket'. In the context of EOW branes, an orientation of the brane forces an $i$ index to contract with a $j$ index, but gauging orientation removes that restriction. For example, $\llangle j_1j_2|i_1i_2\rrangle$ will contain a term proportional to $\delta_{j_1j_2}\delta_{i_1i_2}$. For an unoriented theory, we expect the role of the unitary groups in enumerating representations to be replaced by orthogonal (or symplectic) groups. It would be an interesting exercise to work out the details. Similarly, there may be interesting supersymmetric analogues (corresponding to the classification of \cite{Stanford:2019vob}).

Another simple modification of the model replaces `bosonic' statistics with `fermionic', meaning that we include signs $(-1)^{|\pi|}$ according to whether the permutation is even or odd. This amounts to replacing $q$ in the definition \eqref{eq:IPdef} with its negative, $-q$. We do not know of a particularly strong physical motivation for this, but it is nonetheless an interesting mathematical exercise to see what consequences follow.

From the representation theory perspective we can account for this by altering the auxiliary Hilbert space representation $\tilde{R}$, taking a tensor product with the one-dimensional sign representation of $\Sym(n)$. This has the effect of reflecting the corresponding Young diagram, swapping rows with columns. The upshot is that the state counting formula \eqref{eq:dimreps} is changed only by the constraint on $\lambda$: we take Young diagrams with at most $q$ \emph{columns} (instead of rows). From the Coulomb gas perspective, instead of the constraint of maximal filling $\rho(x)=1$ for $x>\hat{q}$, we exclude any density ($\rho(x)=0$) for $x<-\hat{q}$. The main qualitative difference to the result is that for $n>pq$ there are no representations with nonzero norm, since $U(p)$ representations correspond to diagrams with at most $p$ rows: $\eta=0$ and all states are null.

\subsection{Quantisation of $q$}

Throughout, we took the parameter $q$ to be an integer. But from the gravitational argument of section \ref{ssec:topmot}, it is not at all clear why we should make this restriction. And from the original expression \eqref{eq:IPdef} for the inner product, it is not obvious that integer $q$ should be preferred. Nonetheless, integer $q$ is required if we regard the inner product as resulting from a gauge symmetry acting on additional degrees of freedom (as in section \ref{ssec:gaugemot}), since $q$ is the dimension of the auxiliary Hilbert space.  Furthermore, we used this description to  explain why $\eta$ was positive semidefinite. But in fact, $q\in\ZZ$ is necessary for this to hold: non-integral $q$ leads to negative norms for $n>q+1$.

We can see this by looking at the expression \eqref{eq:etanpq} for the eigenvalues of $\eta$ in a given representation of $\Sym(n)$. From the original form of the inner product, this is guaranteed to be $q^{-n}$ times a polynomial of degree $n$ in $q$, so we can continue to use our result (derived for integer $n$) for any $q$. Null states correspond to zeroes of these polynomials, which are always at integer $q$. If these are simple zeroes (or zeroes of any odd degree), then the norm will become negative for the corresponding representation at some value of $q$. The first case where this occurs is the completely antisymmetric representation (a single column of $n$ boxes, corresponding to $\lambda_i=i-1$ for $i\leq n$ and $\lambda_i=i$ otherwise), for which
\begin{equation}\label{eq:antisymmIP}
	\eta_{p,q}^{(n)}(\lambda) = \frac{q(q-1)\cdots(q-n+1)}{q^n}.
\end{equation}
This becomes negative for $n>q+1$, and in that range some representation always has negative norm. The same negativity was seen in \cite{Marolf:2020xie} from the perspective of the boundary ensemble: the Wishart distribution of the matrix of inner products is a sensible probability distribution with positive measure only for $q>n-1$ ($d>k-1$ in the notation of that paper).

An analogous quantisation result follows from a general set of axioms for a theory of quantum gravity \cite{Colafranceschi:2023urj} (expanding on the framework of \cite{Marolf:2020xie}). In particular, the argument uses positivity of an antisymmetric wavefunction with exponentially many boundaries, generalising the feature that the antisymmetric representation gives the first negative norm states \eqref{eq:antisymmIP}.

The necessity of integer $q$ makes a strong case that quantisation of area (or generalised entropy) should be a general feature of gravity, and for the interpretation of the integer $\exp(\frac{A}{4G_N})$ as the dimension of a Hilbert space of short-distance degrees of freedom (as discussed at the end of section \ref{ssec:gaugemot} and in \cite{Maxfield:2022sio}).

\paragraph{Acknowledgements}

I would like to thank Don Marolf, Steve Shenker and Mark Srednicki for helpful conversations. I am supported by DOE grant DE-SC0021085 and a Bloch fellowship from Q-FARM.

\appendix

\section{The JT inner product}\label{app:JTIP}

In section \ref{ssec:topmot}, we explained how to obtain our inner product from a sum over topologies, giving the concrete example of JT gravity. However, we only explained this at the level of the leading disc topology, and with only the Euler characteristic as an action weighting these configurations. Here we explain how to get the exact inner product, including higher topologies and dynamics of energy.

To do this, we simply skip straight to the answer as given in appendix D of \cite{Penington:2019kki}, which expresses EOW brane states (created with some Euclidean time evolution $\tau$) in terms of energy eigenstates with random wavefunctions:
\begin{equation}\label{eq:EOWwavefunction}
	|\psi_i(\tau)\rangle = \sum_a 2^{\frac{1}{2}-\mu}\Gamma(\mu -\tfrac{1}{2}+i\sqrt{2E_a})e^{-\tau E_a} C_{i,a}|E_a\rangle.
\end{equation}
In the ensemble interpretation, $E_a$ are the eigenvalues of a Hermitian random matrix, which we take to be some fixed numbers (partially fixing the ensemble), while $C_{i,a}$ are independent complex Gaussians with unit variance.

If we work in a narrow microcanonical window, we can regard all the energy factors as being equal and absorb them in a normalisation of the wavefunctions. Having done that, our inner product with $n$ EOW brane states is
\begin{equation}
	\llangle j_1,\ldots j_n|i_1,\ldots,i_n\rrangle = \sum_{a_1,\ldots,a_n} \overline{C_{j_1a_1}^*\ldots C_{j_na_n}^* C_{i_1a_1}\ldots C_{i_1a_1}},
\end{equation}
where the overline indicates expectation value in the Gaussian measure for $C_{ia}$. This is computed by a sum over Wick contractions between $C_{i a}$ and $C_{j a}^*$, which are labelled by permutations $\pi$. This leaves $n-|\pi|$ independent sums over labels $a$, so if they run over $q$ values (as happens if we project the EOW brane states onto a microcanonical window containing $q$ energy eigenstates), we obtain \eqref{eq:IPdef} up to a normalisation factor.

It is not necessary to simplify by ignoring the energy-dependent prefactors. We can include them as explained in section \ref{ssec:realistic} by modifying the state of the auxiliary degrees of freedom, which in this case can be taken to be $n$ copes of the $q$-dimensional Hilbert space spanned by energy eigenstates $|E_a\rangle$ in some range of energies. Corresponding to the factors in \eqref{eq:EOWwavefunction}, we get a density matrix
\begin{equation}
	\tilde{\rho} = \left(\frac{\Gamma(\mu -\tfrac{1}{2}+i\sqrt{2H}) \Gamma(\mu -\tfrac{1}{2}-i\sqrt{2H})}{2^{\mu-1}} e^{-\beta H}\right)^{\otimes n}.
\end{equation}

\section{Representation theory of the inner product}\label{app:reptheoryIP}

Here we expand upon some of the discussion in sections \ref{ssec:gaugemot} and \ref{sec:null}, including a more general inner product as well as various technical details and calculations.

As described in \ref{ssec:gaugemot}, we are interested in an inner product obtained from a pair of vector spaces $\hilb$ and $\tilde{\hilb}$ (in this appendix dropping the $_\mathrm{kin}$ subscripts), which furnish unitary representations $R,\tilde{R}$ of a group $G$. It is straightforward to allow any compact $G$, but we will here restrict to  finite groups to simplify notation. We are interested in an inner product of the form \eqref{eq:Gaverage2},
\begin{equation}\label{eq:etaapp}
	\eta  = \sum_{g\in G}  \widetilde{\Tr}\left(\tilde{\rho}\,\tilde{R}(g)\right) R(g) = |G| \,\widetilde{\Tr} \left(P \tilde{\rho}\right).
\end{equation}
We will take the state $\tilde{\rho}$ on $\tilde{\hilb}$ to be $G$-invariant, meaning that it commutes with $\tilde{R}(g)$ for all elements $g\in G$ of the group.

For the case where $\tilde{\rho}$ is the maximally mixed state, we can write this in terms of the characters $\tilde{\chi}$ of the representation $\tilde{R}$:
\begin{equation}
	\eta  = \frac{1}{\dim \tilde{\hilb}}\sum_{g\in G} \tilde{\chi}(g) R(g) \,.
\end{equation}
More generally, the coefficients of $R(g)$ will be class functions of $G$, meaning that they depend only on the conjugacy class of $g$ (which follows from $G$-invariance of the state). The characters of irreducible representations form a basis of class functions.

To understand $\eta$, as in section \ref{sec:null} we decompose $\hilb$ and $\tilde{\hilb}$ into irreducible representations of $G$. We can write this dscomposition as
\begin{equation}\label{eq:Virreps}
	\hilb= \bigoplus_\lambda U_\lambda \otimes V_\lambda^G, \qquad \tilde{\hilb}= \bigoplus_\lambda \tilde{U}_\lambda \otimes V_\lambda^G,
\end{equation}
where the $\lambda$ in the direct sums labels the inequivalent irreducible representations $R_\lambda$ of $G$ acting on modules $V_\lambda^G$. This means that the matrices of the representations are block diagonal, given by
\begin{equation}
	R(g) = \bigoplus_\lambda \id_{U_\lambda} \otimes R_\lambda(g),\quad \tilde{R}(g) = \bigoplus_\lambda \id_{\tilde{U}_\lambda} \otimes R_\lambda(g).
\end{equation}
 The dimension of $U_\lambda$ and $\tilde{U}_\lambda$ (which can be zero) give the number of times that $\lambda$ appears in the decomposition of $V$ and $\tilde{V}$ respectively into irreducible representations. In the text, $U_\lambda$ and $\tilde{U}_\lambda$ happen to also be irreducible representations of $U(p)$ and $U(q)$, but we only use that fact to give us formulas for their dimensions.
 
Now, since the state $\tilde{\rho}$ is $G$-invariant, by Schur's lemma it is block diagonal in the decomposition \eqref{eq:Virreps} of $\tilde{V}$, and is proportional to the identity in each $V_\lambda^G$. That means that we can characterise $\tilde{\rho}$ by a probability $\mu(\lambda)>0$ for each irrep and a corresponding density matrix $\tilde{\rho}_\lambda$ on $U_\lambda$:
\begin{equation}\label{eq:rhotildeirreps}
	\tilde{\rho} = \bigoplus_\lambda \frac{\mu(\lambda)}{ \dim V_\lambda^G}\, \tilde{\rho}_\lambda \otimes \id_\lambda,
\end{equation}
where $\id_\lambda$ is the identity in $V_\lambda^G$. We define $\mu(\lambda)$ to properly normalise the states as $\Tr(\tilde{\rho}_\lambda)=1$. For example, taking the maximally mixed state $\tilde{\rho}=\frac{1}{\dim\tilde{\hilb}}\id_{\tilde{\hilb}}$ each $\tilde{\rho}_\lambda = \frac{1}{\dim\tilde{U}_\lambda}\id_{\tilde{U}_\lambda}$ is also maximally mixed and $\mu(\lambda) = \frac{\dim\tilde{U}_\lambda \dim V_\lambda^G}{\dim\tilde{\hilb}} $. Normalisation of $\tilde{\rho}$ gives us $\sum_\lambda \mu(\lambda)=1$ in all cases.

Using this decomposition into irreps, we can write the coefficients in the expression \eqref{eq:etaapp} for $\eta$ in terms of characters $\chi_\lambda$ of irreps. Since $\Tr(\tilde{\rho}_\lambda)=1$ (by absorbing the normalisation into $\mu(\lambda)$), we have
\begin{equation}\label{eq:trcharacters}
	  \widetilde{\Tr}\left(\tilde{\rho}\,\tilde{R}(g)\right)= \sum_\lambda \frac{\mu(\lambda)}{ \dim V_\lambda^G} \chi_\lambda(g).
\end{equation}
Putting this together with the decomposition of $V$ gives us
\begin{equation}\label{eq:etadecompsum}
	\eta  =  \bigoplus_\lambda \id_{U_\lambda} \otimes \left(\sum_{g\in G} R_\lambda(g) \sum_{\lambda'} \frac{\mu(\lambda')}{ \dim V_{\lambda'}^G} \chi_{\lambda'}(g)\right).
\end{equation}

With this expression we can evaluate the sum over group elements $g$ using Schur orthogonality, a standard result from representation theory. This states that for a finite (or more generally compact) group $G$, the matrix elements of unitary irreducible representations form an orthogonal basis of functions on $G$, under the inner product defined by averaging over group elements. In equations, if $(R_\lambda(g))_{mn}$ are the matrix elements of the representation $V_\lambda^G$ (with $m,n=1,\ldots,\dim V_\lambda^G$), we have
\begin{equation}
	\frac{1}{|G|}\sum_g \left(R_\lambda(g)\right)_{m n}\overline{\left(R_{\lambda'}(g)\right)}_{m' n'} = \frac{1}{\dim V_\lambda^G} \delta_{\lambda\lambda'}\delta_{mm'}\delta_{nn'},
\end{equation}
which applies in any basis for the representations in which all the $R_\lambda(g)$ are unitary. We need a result with a sum over characters, obtained by setting $m'=n'$ and summing over their values:
\begin{equation}\label{eq:chirhoOrthog}
	\frac{1}{|G|}\sum_g R_\lambda(g)\overline{\chi_{\lambda'}(g)} = \frac{\delta_{\lambda\lambda'}}{\dim V_\lambda^G}  \id_{V_\lambda^G}.
\end{equation}

This is almost precisely the sum over $G$ appearing in \eqref{eq:etadecompsum}. The only difference is the complex conjugate of the character, which we can absorb into the label $\lambda'$ for the representation (swapping each irrep $\lambda'$ with its conjugate representation $\bar{\lambda}'$). The orthogonality picks out the term $\lambda'=\bar{\lambda}$ in the sum \eqref{eq:etadecompsum}, giving us
\begin{equation}
	\eta  =  \bigoplus_\lambda \mu(\bar{\lambda})  \frac{|G|}{(\dim V_\lambda^G)^2} \id_{U_\lambda} \otimes     \id_{V_\lambda^G}.
\end{equation}
For the special case of a maximally mixed state $\tilde{\rho}$, this gives us
\begin{equation}
	\eta  =  \frac{|G|}{\dim\tilde{\hilb}}\bigoplus_\lambda \frac{\dim\tilde{U}_\lambda }{ \dim V_\lambda^G }   \id_{U_\lambda} \otimes     \id_{V_\lambda^G}.
\end{equation}
This gives \eqref{eq:etalambda} in the main text.
From these expressions, a few properties are manifest. First, $\eta$ is positive semi-definite. Second, the kernel of $\eta$ corresponds to terms with $\mu(\bar{\lambda})=0$ (and $\dim U_\lambda>0$ so such states appear in $\hilb$ in the first place). That is, the null states are precisely those in a representation $\lambda$ whose dual $\bar{\lambda}$ does not appear in the decomposition \eqref{eq:rhotildeirreps} of $\tilde{\rho}$ into irreps. For the maximally mixed state, this happens when $\dim\tilde{U}_\lambda$ is zero (while $\dim U_\lambda$ is nonzero).

For a nice special case we can choose $\tilde{\hilb}$ to be the regular representation of $G$ and $\tilde{\rho}$ maximally mixed. Then $\mu(\lambda)$ becomes the Plancherel measure $\mu(\lambda)=\frac{(\dim V_\lambda^G)^2}{|G|}$, and $\eta$ is simply the identity.

In particular, from this result we can get the rank of $\eta$, which is the dimension of the physical Hilbert space. For this, $\tilde{\hilb}$ and $\tilde{\rho}$ are relevant only for identifying which $\lambda$ are absent (with $\mu(\lambda)=0$), and we have
\begin{equation}
	\rank \eta = \sum_{\substack {\lambda\\ \mu(\lambda)\neq 0}} \dim U_\lambda \dim V_\lambda^G\,.
\end{equation}
In the case studied in the text, this is the main result \eqref{eq:dimreps} of section \ref{sec:null} giving the dimension in terms of sums over Young diagrams with a bounded number of rows.

\section{Coulomb gas calculations}\label{app:Coulomb}

\subsection{Minimising a Coulomb gas energy}

We would like to solve the problem of minimising a Coulomb gas Hamiltonian \eqref{eq:dimHam2} subject to inequality constraints \eqref{eq:rhoineq} and equality constraints \eqref{eq:rhoConstraint} and \eqref{eq:dipole}. This is amenable to some standard methods \cite{Eynard:2015aea,Saad:2019lba}.

First, we are guaranteed a unique local minimum because the Hamiltonian $\hat{\mathsf{H}}$ is  quadratic in $\rho$ with a positive-definite quadratic term, and the constraints (both equalities and inequalities) are linear in $\rho$.  To see that the Coulomb interaction term is a positive quadratic form, express it in terms of the Fourier transform $\hat{\rho}(k)=\int dx\, e^{ik x}\rho(x)$:
\begin{equation}
	- \int \log|x_1-x_2| \delta\rho(x_1)\delta\rho(x_2) dx_1 dx_2 = \int \frac{|\delta\hat{\rho}(k)|^2}{2|k|}dk.
\end{equation}
This integral converges for variations $\delta\rho$ which solve the constraints \eqref{eq:rhoConstraint} and \eqref{eq:dipole}, since in Fourier space they become $\delta\hat{\rho}(0)= \delta\hat{\rho}'(0) =0$ (and since $\rho(x)$ is compactly supported, $\hat{\rho}(k)$ is smooth).

To find the minimum we vary the density, and the corresponding variation of the energy is given by an effective potential $\mathsf{V}_\mathrm{eff}(x)$:
\begin{equation}
\begin{aligned}
	\delta \hat{\mathsf{H}} &= \int \mathsf{V}_\mathrm{eff}(x)\delta\rho(x) dx \\
	\mathsf{V}_\mathrm{eff}(x) &= \hat{\mathsf{V}}(x) -2\int \log|x-y|\rho(y) dy + \nu +\mu x,
\end{aligned}
\end{equation}
where $\nu$ and $\mu$ are Lagrange multipliers which impose the constraints \eqref{eq:rhoConstraint} and \eqref{eq:dipole} respectively. To be at the minimum, we need $\mathsf{V}_\mathrm{eff}(x)=0$ in regions where inequalities are not saturated ($0<\rho(x)<1$), $\mathsf{V}_\mathrm{eff}(x)>0$ in regions where $\rho(x)=0$, and $\mathsf{V}_\mathrm{eff}(x)<0$ where $\rho(x)=1$.

In regions where inequalities are not saturated, we get the natural condition that the `force' on the particles $-\mathsf{V}_\mathrm{eff}'(x)$  vanishes (and this also eliminates dependence on the Lagrange multiplier $\nu$). This gives us
\begin{equation}\label{eq:eqbm}
	2\mathcal{P}\int \frac{\rho(y)}{x-y} dy = \hat{\mathsf{V}}'(x) + \mu,
\end{equation}
where $\mathcal{P}\int$ is the Cauchy principal value integral. For an algorithmic approach to construct the solution $\rho$ from $\hat{\mathsf{V}}$, it's helpful to express this relation in terms of the resolvent $R(z)$, defined by
\begin{equation}\label{eq:resolvent}
	R(z) = \int \frac{\rho(x)}{z-x} dx
\end{equation}
for $z\in \CC - \supp\rho$, where $\supp\rho$ is the support of $\rho$ (the region of the real line where $\rho\neq 0$). This is analytic away from $\supp\rho$, and $\rho$ is determined by its discontinuity:
\begin{equation}\label{eq:rhodisc}
	\rho(x) = -\frac{1}{2\pi i}(R(x+i\epsilon)-R(x-i\epsilon)).
\end{equation}
On the other hand, the principal value integral appearing in the equilibrium relation \eqref{eq:eqbm} is given by the average of the resolvent on the two sides of the cut, so we find
\begin{equation}\label{eq:eqbmRes}
	\hat{\mathsf{V}}'(x)+\mu = R(x+i\epsilon)+R(x-i\epsilon).
\end{equation}

Now we can compute the resolvent $R(z)$ using a trick  \cite{Eynard:2015aea}. This first requires us to assume that $\rho$ is nonzero on a single interval $[x_-,x_+]$, and that $\rho$ never saturates its upper bound. We can check consistency a posteriori (since there is a unique local minimum). These assumptions will in fact always hold in the interesting regime that $\hat{q}$ is small enough for the constraint on the number of rows to be important. These assumptions ensure that $R'(z)$ is analytic on $\CC-[x_-,x_+]$, and that \eqref{eq:eqbmRes} holds for $x\in [x_-,x_+]$. Using these properties, we can write
\begin{align}
	R(z) &= \oint_z \frac{dw}{2\pi i} \frac{R(w)}{w-z} \sqrt{\frac{(z-x_-)(z-x_+)}{(w-x_-)(w-x_+)}} \\
	&= -\oint_C \frac{dw}{2\pi i} \frac{R(w)}{w-z} \sqrt{\frac{(z-x_-)(z-x_+)}{(w-x_-)(w-x_+)}} \\
	&=  \int_{x_-}^{x_+}\frac{\hat{\mathsf{V}}'(x)+\mu}{2\pi (z-x)}\sqrt{\frac{(z-x_-)(z-x_+)}{(x-x_-)(x_+-x)}}dx \,.\label{eq:Rintegral} 
\end{align}
For the first line we take a small contour where $w$ encircles $z$, which in the second line is deformed to the contour $C$ encircling the interval $[x_-,x_+]$, using the fact that $R'(w)$ is analytic away from the interval and decays like $\frac{1}{w^2}$ as $w\to\infty$ (see \eqref{eq:resolventExpansion}). In the third line we use the fact that $\sqrt{(w-x_-)(w-x_+)}$ takes opposite signs above and below the interval, so when they are combined we get the sum of $R'$ on either side of the cut, which we replace with $\mathsf{V}''$ using \eqref{eq:eqbmRes}.

The integral \eqref{eq:Rintegral} does not quite determine $R$, since it depends on three unknown free parameters: then chemical potential $\mu$, and the endpoints $x_\pm$ of the interval. We can fix these  using the large $z$ expansion of the resolvent, which is the multipole expansion of $\rho$ determined by writing the denominator of \eqref{eq:resolvent} as a geometric series:
\begin{equation}\label{eq:resolventExpansion}
\begin{aligned}
	R(z) &\sim z^{-1} \int \rho(x) dx + z^{-2} \int x \rho(x) dx + \cdots \\
	 &= \hat{q}z^{-1} + \left(\tfrac{1}{2}\hat{q}^2-\hat{n}\right) z^{-2} + \cdots.
\end{aligned}	
\end{equation}
This gives us three equations (from the $z^0$, $z^{-1}$ and $z^{-2}$ coefficients) for the three unknowns. So after we've evaluated \eqref{eq:Rintegral} to find $R(z)$, we may use its expansion at large $z$ to fix the parameters.

In particular it is very straightforward to determine $\mu$, since the $\mu$-dependent term in \eqref{eq:Rintegral} evaluates to a constant $\frac{\mu}{2}$. Hence, we simply choose $\mu$ to cancel the constant term in the large $z$ expansion \eqref{eq:resolventExpansion}. It then remains only to determine $x_\pm$ from \eqref{eq:resolventExpansion}, and to check a posteriori that our solution obeys the inequality constraints $0\leq \rho(x)\leq 1$.

\subsection{Finding the density}

We now implement the above strategy for our potential \eqref{eq:Vhat}. We can write this as the sum of two terms, each giving a logarithmic force:
\begin{equation}
\begin{gathered}
	\hat{\mathsf{V}}(x) =2 \mathsf{V}_{\hat{q}}(x)- \mathsf{V}_{\hat{p}}(x),\\
	\mathsf{V}_X(x)= (X-x)(\log(X-x)-1)\implies - \mathsf{V}_X'(x) = \log(X-x) \,.
\end{gathered}
\end{equation}
Hence, our main job is to calculate the contribution of such a potential to the resolvent from the integral \eqref{eq:Rintegral}:
\begin{equation}
	R_{X}(z) =-\int_{x_-}^{x_+}\frac{\log(X-x)}{2\pi (z-x)}\sqrt{\frac{(z-x_-)(z-x_+)}{(x-x_-)(x_+-x)}}dx.
\end{equation}
This will give us
\begin{equation}
	R(z) = 2R_{\hat{q}}(x)- R_{\hat{p}}(x) + \frac{\mu}{2}.
\end{equation}

%
%

To evaluate this integral, first differentiate with respect to $X$. That removes the logarithm and gives us something easier to integrate:
\begin{equation}
\begin{aligned}
	\partial_X R_{ X}(z) &= -\int_{x_-}^{x_+}\frac{1}{2\pi (z-x)(X-x)}\sqrt{\frac{(z-x_-)(z-x_+)}{(x-x_-)(x_+-x)}}dx \\
	&= \frac{1}{2}\frac{1}{X-z}\left(\sqrt{\frac{(z-x_-) (z-x_+)}{(X-x_-) (X-x_+)}}-1\right).
\end{aligned}	
\end{equation}
To compute $R_X(z)$ itself we need to integrate this with respect to $X$, and to fix the integration constant we can match to the large $X$ expansion
\begin{equation}
	R_X(z) \sim -\tfrac{1}{2}\log X +O(X^{-1}) \qquad (X\to\infty).
\end{equation}
The result is
\begin{equation}
	R_X(z) = \frac{1}{2}\log\left(\frac{1}{X-z}\frac{\sqrt{\frac{z-x_+}{z-x_-}}+1}{\sqrt{\frac{z-x_+}{z-x_-}}-1}\frac{\sqrt{\frac{(X-x_-) (z-x_+)}{(X-x_+)(z-x_-)}}-1}{\sqrt{\frac{(X-x_-) (z-x_+)}{(X-x_+)(z-x_-)}}+1}\right).
\end{equation}
As required, this is analytic away from a branch cut running from $z=x_-$ to $z=x_+$ (in particular, despite appearances $R_X$ is perfectly analytic at $z=X$ since a zero in the last factor cancels the pole in the first).


From $R_X(z)$ we  determine the corresponding density $\rho_X(x)$ from the discontinuity across the branch cut running between $x_-$ and $x_+$, as in \eqref{eq:rhodisc}:
\begin{equation}
	\rho_X(x) = \frac{i}{2\pi}\log\left(\frac{1+i\sqrt{\frac{x_+-x}{x-x_-}}}{1-i\sqrt{\frac{x_+-x}{x-x_-}}}\frac{1-i\sqrt{\frac{(X-x_-) (x_+-x)}{(X-x_+)(x-x_-)}}}{1+i\sqrt{\frac{(X-x_-) (x_+-x)}{(X-x_+)(z-x_-)}}}\right).
\end{equation}
More simply, this can be written as
\begin{equation}
\begin{gathered}
	\rho_X(x) = \frac{1}{2\pi}\sin^{-1} f_\infty(x)-\frac{1}{2\pi}\sin^{-1}f_X(x),
\end{gathered}
\end{equation}
where $f_X$ is the fractional linear (M\"obius) map for which $f_X(x_\pm)=\pm 1$ and $f_X(X)=\infty$:
\begin{equation}\label{eq:fX}
	f_X(x) = \frac{1}{x_+-x_-} \left(\frac{2 (X - x_-) (X - x_+)}{X - x} + x_+ + x_- - 2 X \right), \quad f_\infty(x) = \frac{2x-x_--x_+}{x_+-x_-}.
\end{equation}
With this we can write the solution for our potential by adding the contribution from the two terms $\mathsf{V}(x) = 2\mathsf{V}_{\hat{q}}(x)-\mathsf{V}_{\hat{p}}(x)$ (the constant from the Lagrange multiplier $\mu$ does not contribute to $\rho$):
\begin{equation}
\begin{aligned}
	\rho(x) &= 2\rho_{\hat{q}}(x)-\rho_{\hat{p}}(x) \\
	&= \frac{1}{2\pi}\sin^{-1} f_\infty(x)-\frac{1}{\pi}\sin^{-1}f_{\hat{q}}(x)+\frac{1}{2\pi}\sin^{-1}f_{\hat{p}}(x).
\end{aligned}
\end{equation}
This gives us an admissible density function satisfying the inequality constraints $0\leq \rho(x)\leq 1$ for any values of parameters $x_-<x_+<\hat{q}<\hat{p}$. To see that $\rho \leq 1$ we simply observe that $\sin^{-1}$ takes values between $-\frac{\pi}{2}$ and $\frac{\pi}{2}$ (and saturation can only occur for $x=x_+$ when $x_+ = \hat{q}$). For positivity, first note how this function behaves near the endpoints: it vanishes as a square root ($\sqrt{x-x_-}$ or $\sqrt{x_+-x}$) with positive coefficient. So $\rho(x)$ can only become negative if it has at least three stationary points. But $\rho'(x)=0$ requires that $x$ solves a quadratic equation in $x$, which has at most two solutions (and given endpoint behaviour, there must be exactly one solution in the interval $x_-<x<x_+$).

However, our result for $\rho(x)$ depends on the endpoints $x_\pm$ of the interval. Our next task is therefore to understand these parameters in terms of $\hat{p}$, $\hat{q}$ and $\hat{n}$.

\subsection{Fixing the parameters}

To determine $x_\pm$, we use the constraints on the integrated density and its dipole moment. These are conveniently determined from the large $z$ expansion of the resolvent as in \eqref{eq:resolventExpansion}.

For $R_X(z)$ above, the coefficients in this expansion are given by
\begin{equation}
\begin{gathered}
R_X(z) \sim R_X^{(0)} + R_X^{(1)} z^{-1} + R_X^{(2)} z^{-2}+ \ldots \\
R_X^{(0)} = \frac{1}{2}\log\left(\frac{4}{x_+-x_-}\frac{\sqrt{\frac{X-x_-}{X-x_+}}-1}{\sqrt{\frac{X-x_-}{X-x_+}}+1}\right),  \\
R_X^{(1)} = \frac{1}{4} (2 X-x_--x_+)-\frac{1}{2} \sqrt{(X-x_-) (X-x_+)},  \\
R_X^{(2)} = \frac{1}{32} \left(8 X^2-3 x_-^2-2 x_- x_+-3 x_+^2\right)-\frac{1}{8} (2 X+x_-+x_+) \sqrt{(X-x_-) (X-x_+)}.
\end{gathered}
\end{equation}
 Our constraints therefore become
\begin{align}
	2R_{\hat{q}}^{(1)} -R_{\hat{p}}^{(1)} &= \hat{q},\label{eq:constr1} \\
	2R_{\hat{q}}^{(2)} -R_{\hat{p}}^{(2)} &= \tfrac{1}{2}\hat{q}^2-\hat{n} \,. \label{eq:constr2}
\end{align}
The chemical potential $\mu$ (which gives the derivative of the minimal energy with respect to $\hat{n}$) is given by
\begin{equation}
	\mu = 2 R_{\hat{p}}^{(0)} - 4 R_{\hat{q}}^{(0)}.
\end{equation}

These are rather complicated non-linear equations so we do not attempt to give explicit general solutions for $x_\pm$ in terms of $\hat{q},\hat{p},\hat{n}$. We note only that there is nontrivial dependence on just two of these parameters, since the problem has a symmetry under simultaneous rescaling of $x$, $\hat{p}$, $\hat{q}$ and $\sqrt{\hat{n}}$. In the text, we describe the solutions in various limits.

\section{Connection to the longest increasing subsequence problem}\label{app:LIS}

Given a permutation $\pi\in \Sym(n)$, a subsequence $i_1<i_2<\ldots<i_k$ is increasing if $\pi(i_1)<\pi(i_2)<\ldots<\pi(i_k)$. For any $pi$, let $L(\pi)$ be the largest $k$ for which a sequence exists, i.e.~the length of the longest increasing subsequence (an integer between $1$ and $n$). If we select a permutation $\pi$ uniformly at random from $\Sym(n)$, what does the distribution of $L(\pi)$ looks like? This `longest increasing subsequence' problem was first suggested by Ulam \cite{ulam1961monte} as an example of a problem amenable to Monte-Carlo methods in 1961.

It turns out that our dimension counting problem for large $p$ ($p\gg q$ and $p\gg\sqrt{n}$) can be restated in these terms.  The proportion of physical states $\frac{d^{(n)}_{p,q}}{p^n}$ is precisely equal to the probability that  $L(\pi)\leq q$. And indeed, the first solution to the longest increasing sequence problem is essentially equivalent to out Coulomb gas calculations \cite{vershik1985asymptotic,logan1977variational}.

The connection passes through the Robinson-Schensted algorithm, which associates a Young diagram $\lambda$ with $n$ boxes to any permutation $\pi \in \Sym(n)$. The number of different permutations $\pi$ leading to any given $\lambda$ is precisely $(\dim V^{\Sym(n)}_\lambda)^2$, so selecting a uniformly random permutation leads to a Young diagram selected from the Plancherel measure. The length of the longest increasing subsequence in $\pi$ is given by the length of the first row of $\lambda$, and the length of the longest decreasing subsequence is similarly the height of the first column, or equivalently the number of rows of $\lambda$. For a review of the details and much more see \cite{romik2015surprising}.

\bibliographystyle{JHEP}

\bibliography{PSNullStates,mathbib}

\end{document}